\documentclass[twocolumn,trackchanges]{aastex7}
\usepackage{txfonts}

\begin{document}

\title{Absolute Calibration of Cluster Mira Variables to Provide a New Anchor for the Hubble Constant Determination}


\author[orcid=0000-0001-6147-3360,sname='Bhardwaj']{Anupam Bhardwaj}
\affiliation{Inter-University Centre for Astronomy and Astrophysics (IUCAA), Post Bag 4, Ganeshkhind, Pune 411 007, India}
\email[show]{anupam.bhardwaj@iucaa.in}  

\author[]{Noriyuki Matsunaga}
\affiliation{Department of Astronomy, School of Science, The University of Tokyo, 7-3-1 Hongo, Bunkyo-ku, Tokyo 113-0033, Japan}
\email{matsunaga@astron.s.u-tokyo.ac.jp }

\author[orcid=0000-0001-6169-8586]{Caroline D. Huang}
\affiliation{Center for Astrophysics, Harvard \& Smithsonian, 60 Garden Street, Cambridge, MA 02138, USA}
\email{caroline.huang@cfa.harvard.edu}

\author[orcid=0000-0002-6124-1196]{Adam G. Riess}
\affiliation{Space Telescope Science Institute, 3700 San Martin Drive, Baltimore, MD 21218, USA}
\affiliation{Department of Physics and Astronomy, Johns Hopkins University, Baltimore, MD 21218, USA}
\email{ariess@stsci.edu}

\author[orcid=0000-0002-6577-2787]{Marina Rejkuba} 
\affiliation{European Southern Observatory, Karl-Schwarzschild-Stra\ss e 2, 85748, Garching, Germany}
\email{mrejkuba@eso.org}


\begin{abstract}

Mira variables in globular clusters can provide an accurate and precise absolute calibration of their period-luminosity relations (PLRs) to independently anchor the cosmic distance scale and determine the Hubble constant. We present homogeneous near-infrared ($JHK_s$) time-series photometric observations of a sample of 55 candidate long-period variables in 18 globular clusters covering a wide metallicity range ($-1.7 < \textrm{[Fe/H]} < -0.1$ dex). The Gaia proper motions, long-period variability information, and optical-infrared colors are used to identify 41 oxygen-rich Miras as members of the globular clusters. Mean luminosities of Miras in the $JHK_s$ bands are independently calibrated using the recommended distances and mean parallaxes to their host clusters. Cluster Mira PLRs exhibit scatter comparable to the Large Magellanic Cloud (LMC) variables and do not show any dependence on iron abundance for a wide range of metallicities. We establish the accuracy of cluster Miras as independent anchors by determining a distance modulus to the LMC, $18.45 \pm 0.04$ mag, in agreement with the 1.2\% precise geometric distance. Our $H$-band photometry is transformed to derive Hubble Space Telescope F160W PLR for cluster Miras providing a three-anchor baseline with the LMC and NGC 4258. We employ three-anchor solution to determine distances to two type Ia supernovae host galaxies, NGC 1559 ($31.39\pm0.05$ mag) and M101 ($29.07\pm0.04$ mag), and provide a $3.7\%$ measurement of the Hubble constant, $H_0 = 73.06\pm 2.67$ km~s$^{-1}$~Mpc$^{-1}$. Similar to Cepheids, our independent baseline solution results in a local $H_0$ determination that is systematically larger than its inference from the early universe probes, further supporting the ongoing Hubble tension.
\end{abstract}

\keywords{\uat{Mira variable stars}{1066} --- \uat{Distance indicators}{394} --- \uat{Globular star clusters}{656} --- \uat{Cosmology}{343} --- \uat{Distance Scale}{394} --- \uat{Hubble constant }{758}}

\section{Introduction} \label{sec:intro}
Pulsating variable stars such as Cepheids, RR Lyrae, and Miras serve as primary standard candles for the extragalactic distance ladder leading to the determination of the Hubble constant ($H_0$) -- the present expansion rate of the Universe \citep{freedman2001, riess2022, bhardwaj2023a, huang2024, riess2024a}. The most precise direct measurements of $H_0$ based on classical Cepheids and type Ia supernovae (SNe Ia) are currently in significant discord with its model-dependent inference based on cosmic microwave background observations from the Planck mission \citep{planck2020, riess2022}. This `Hubble Tension' is at the level of $> 5\sigma$ such that local $H_0$ determinations are significantly larger than the predictions of $\Lambda$CDM, possibly hinting at new physics in the standard cosmological model \citep{valentino2021, abdalla2022, hu2023}. Cepheid-independent calibrations of the distance ladder based on stellar standard candles with different ages and metallicities, for example the tip of the red giant branch (TRGB), J-region asymptotic giant branch (J-AGB), and Mira stars, are being developed to better evaluate the significance of the Hubble tension \citep{freedman2024, huang2024, li2025}. 

Mira variables are low to intermediate-mass stars at the top of the asymptotic giant branch (AGB) evolution, where they pulsate with typical periods between 100 and 1000 d \citep{catelan2015}. Miras exhibit large visual band amplitudes ($> 2.5$ in $V$-band) and their typical classification in the near-infrared (NIR) bands requires regular near-sinusoidal light curves with $\Delta K_s>0.4$ mag \citep{matsunaga2017}. These are the most well-known members of an entire family of long-period variables (LPVs) which includes small amplitude red giants to red supergiant stars \citep{wood2015, mowlavi2018}. Miras oscillate solely in the fundamental mode displaying near-sinusoidal long-term light curves with quasi-periodic changes in the period and amplitudes. Like other AGB stars, Miras are separated into oxygen-rich (O-rich) and carbon-rich (C-rich) cool giants. Their surface chemistry is determined by the dredge up episodes by the deep convective envelope during nucleosynthesis which can bring carbon produced in the interior to the surface \citep{riebel2010, hofner2018}. The circumstellar extinction significantly impacts the properties of long-period C-rich Mira stars and they appear fainter than the PLRs for O-rich Miras at optical and NIR wavelengths \citep{ita2011, yuan2017b}. 

From the earliest studies on NIR observations of Miras \cite[e.g.,][]{glass1981, menzies1985, feast1989}, it was well known that these variables follow a tight period-luminosity relation (PLR) with a typical scatter of $\sim0.25$ mag using single or a few epochs of measurements. Later, the studies of LPVs in microlensing surveys led to the discovery of multiple PLRs for giants \citep{wood2000, cioni2001, ita2004a}. With modern time-domain photometric surveys and infrared detectors, Mira PLRs, in particular in the Large Magellanic Cloud (LMC), have been obtained with accuracy and precision comparable to the classical Cepheids in the $K_s$ band \citep[$\sigma\lesssim 0.15$~mag,][]{ita2011, yuan2017b}. This has led to Miras being calibrated at a high enough precision to determine Mira-based distances to SNe Ia host galaxies providing an independent $H_0$ determination \citep{huang2024}. 

Currently, Miras in the LMC and maser-host galaxy NGC 4258 play the crucial role of calibrating their PLR using geometric distances to these galaxies. The absolute calibration of Miras in our own Galaxy is difficult due to uncertainties associated with parallaxes of these bright, red stars with extended atmospheres and large amplitude variations, and lack of homogeneous reddening and metallicity estimates. \cite{sanders2023} used Gaia data release 3 (DR3) parallaxes of Milky Way field variables to calibrate the Mira distance scale. However, large parallax uncertainties coupled with random-epoch NIR observations from Two Micron All Sky Survey \citep[2MASS,][]{skrutskie2006} result in large scatter in the Mira PLR, making those unreliable for a precise $H_0$ determination \citep[also see discussion in][]{huang2024}. Mira variables in star clusters can potentially provide a more accurate and precise calibration of their PLRs due to availability of independent distances and improvements in the mean parallax from thousands of cluster member stars \citep[e.g. for Cepheids in][]{riess2022a}. This work presents homogeneous time-series photometry of Mira variables in globular clusters and uses their luminosity calibrations to determine a precise baseline value of $H_0$ together with LMC and NGC 4258 as independent anchors.

The manuscript presents description of photometric data, light curves of LPVs, and cluster properties in Section~\ref{sec:data}, and the calibration of Mira PLRs in Section~\ref{sec:lpv_plr}. The $H_0$ determination is described in Section~\ref{sec:h0} and the results are discussed in Section~\ref{sec:discuss}.

\section{The Data} \label{sec:data}

\begin{figure*}
\centering
\includegraphics[width=0.99\textwidth]{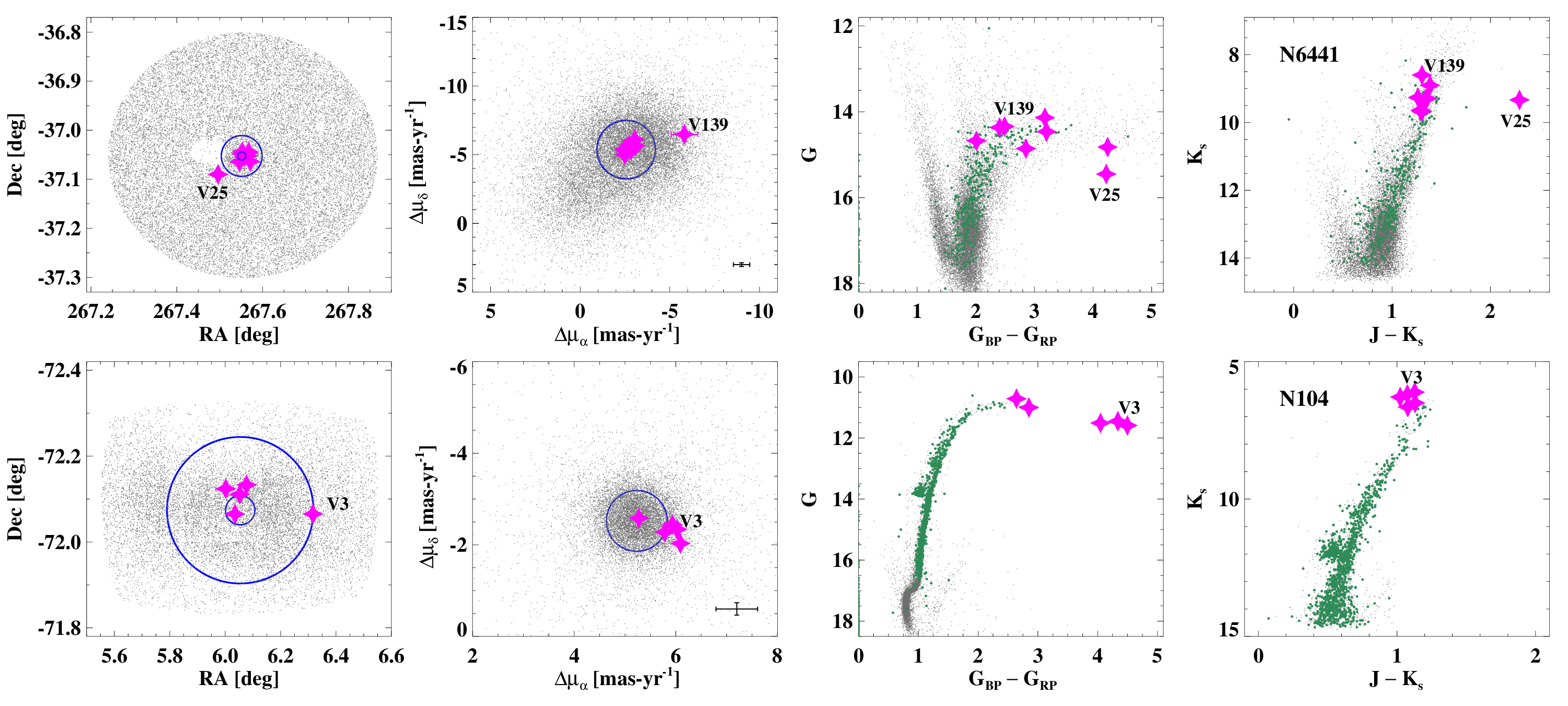}
\caption{The spatial distribution, proper motions, and optical-NIR color-magnitude diagrams of LPVs in N6441 (top) and N104 (bottom). The small/big circles in the coordinates represent 1/5 half light radius \citep[$r_h\textrm{(N6441)}=0.57\arcmin$ and $r_h\textrm{(N104)}=3.17\arcmin$,][]{harris2010}. The circles in the proper motion planes represent average scatter around the mean values along both axes. Representative $\pm5\sigma$ errors in the proper motions are also shown. In the color-magnitude diagrams, green circles represent stars that are common within 5 half light radius and $1\sigma$ scatter in the proper motions. LPVs in their host clusters are shown with magenta stars. The optical-NIR magnitudes are not corrected for extinction.}
\label{fig:spatial}
\end{figure*}

Miras exhibit relatively large amplitudes even at longer wavelengths \citep[$\Delta K_s > 0.4$ mag,][]{whitelock2008, matsunaga2017}, and therefore time-series observations are crucial to derive accurate mean magnitudes and PLRs. We utilized homogeneous time-series observations of 55 candidate LPVs obtained as part of a monitoring program of globular clusters at South African Astronomical Observatory \citep[SAAO,][]{matsunaga2006, matsunaga2007}. These observations were obtained with a cadence of 40-60 days in $JHK_s$ filters between 2002 and 2005, and were collected using the SIRIUS (Simultaneous-color InfraRed Imager for Unbiased Surveys) NIR camera mounted at the 1.4 m Infrared Survey Facility (IRSF) telescope. The photometric data reduction was carried out using the DoPHOT software \citep{matsunaga2007}, and the light curves were calibrated in 2MASS system \citep{skrutskie2006}. Thirty two of these LPV candidates were previously utilized to probe the mass-loss and dust production in evolved stars in globular clusters \citep{sloan2010}. 

To obtain accurate periods, we extended the time baseline of NIR observations of our cluster LPVs by compiling additional data from \cite{frogel1981} for four NGC 104 stars and from \cite{menzies1985} for 16 variables. These literature datasets were homogenized by \citet{feast2002} in the SAAO system, and were transformed to the 2MASS system using transformations from \citet{koen2007}. 

The sample of 55 IRSF LPVs was crossmatched with the Gaia DR3 astrometry and the LPV catalogs \citep{vallenari2023, lebzelter2023}. While all IRSF LPVs have Gaia astrometry, only 44 (of 55) variables are listed in the Gaia LPV catalog, for which we compile preliminary periods, variability amplitude, and C-type star flag, if available. The subclassification of all 55 IRSF LPVs in Mira (M), semi-regular (SR) variables was obtained from the \citet{clement2001} catalog of variable stars, and uncertain classifications were flagged as M/SR subtype.

Fig.~\ref{fig:spatial} displays spatial distribution, proper motions, and optical-NIR color-magnitude diagrams of LPVs in our sample in two representative clusters, N6441 and N104. N6441 has the largest number of LPVs in our sample (Ter 5 also has 8 LPVs) and is located in the bulge with relatively high reddening and crowding. N104 represents the cluster with the second largest number of five LPVs, but has least reddening in the Globular Cluster (GC) sample. All sources that are within $15\arcmin$ from the cluster center and have both Gaia and 2MASS photometry are shown. These properties of LPVs displayed in Fig.~\ref{fig:spatial} for all clusters are used in Section~\ref{sec:lpv_plr} to identify cluster members and O-rich Mira variables.

\begin{deluxetable}{cccccccccccccccc}
\tablecaption{Pulsation properties of 55 LPVs in globular clusters in the homogeneous IRSF sample. \label{tbl:lpv_data}}
\tabletypesize{\footnotesize}
\tablehead{
{Cluster} & {Variable ID} & {$P$ (days)} & {Class} & \multicolumn{3}{c}{magnitudes} & \multicolumn{3}{c}{$\sigma_\textrm{mag}$} & \multicolumn{3}{c}{amplitudes} & Flag & Remarks\\
      &      &       &       &   $J$ &   $H$ &   $K_s$ & $J$ &   $H$ &   $K_s$ & $J$ &   $H$ &   $K_s$ & &
}
\startdata
    IC1276& V1&    221.0&      M&     8.42&     7.30&     6.76&     0.03&     0.03&     0.03&     0.43&     0.38&     0.35&      Y  &    O-rich, member \\
    IC1276& V3&    304.5&      M&     8.20&     7.14&     6.52&     0.04&     0.04&     0.04&     0.87&     0.95&     0.93&      Y  &    O-rich, member \\
    Lynga7& V1&    542.6&      M&    11.25&     9.01&     7.18&     0.05&     0.05&     0.05&     1.19&     1.28&     1.23&      N  &   C-rich  \\
      N104& V1&    211.7&   M/SR&     7.24&     6.50&     6.11&     0.03&     0.03&     0.03&     0.66&     0.67&     0.66&      Y &    O-rich, member\\
      N104& V2&    198.7&      M&     7.27&     6.58&     6.19&     0.03&     0.03&     0.04&     0.78&     0.76&     0.74&      Y &    O-rich, member\\
      N104& V3&    191.9&      M&     7.31&     6.62&     6.29&     0.03&     0.03&     0.03&     0.64&     0.68&     0.67&      Y &    O-rich, member\\
      N104& V4&    162.9&      M&     7.64&     6.87&     6.50&     0.03&     0.03&     0.03&     0.53&     0.53&     0.50&      Y &    O-rich, member\\
      N104& V8&    151.8&      M&     7.73&     6.97&     6.65&     0.03&     0.03&     0.03&     0.44&     0.38&     0.38&      Y &    O-rich, member\\
     N362& V16&    138.6&      M&     9.13&     8.42&     8.21&     0.03&     0.03&     0.03&     0.62&     0.64&     0.62&      Y  &   O-rich, member\\
      N362& V2&    106.6&   M/SR&     9.74&     9.09&     8.86&     0.02&     0.03&     0.03&     0.31&     0.36&     0.32&      Y &    O-rich\\
     N5139& V2&    236.8&      M&     7.35&     6.69&     6.20&     0.04&     0.04&     0.04&     0.89&     0.80&     0.74&      N &    Field variable, beyond $5r_h$\\
    N5139& V42&    147.6&   M/SR&     8.26&     7.52&     7.16&     0.03&     0.03&     0.03&     0.77&     0.74&     0.70&      Y &    O-rich, member\\
     N5927& V1&    202.8&   M/SR&     9.17&     8.27&     7.86&     0.03&     0.03&     0.03&     0.51&     0.51&     0.50&      Y &    O-rich, member\\
     N5927& V3&    300.0&      M&     8.69&     7.81&     7.28&     0.04&     0.04&     0.04&     0.98&     0.97&     0.94&      Y &    O-rich, member\\
    N6352& V4&    155.0&     SR&     8.58&     7.55&     7.32&     0.02&     0.02&     0.03&     0.16&     0.09&     0.09&      N & small amplitude variable\\
     N6352& V5&    177.0&      M&     8.37&     7.44&     7.05&     0.02&     0.03&     0.03&     0.40&     0.41&     0.40&      Y &    O-rich, member\\
     N6356& V1&    227.0&      M&    10.08&     9.28&     8.80&     0.03&     0.03&     0.03&     0.62&     0.64&     0.62&      Y &    O-rich, member\\
     N6356& V3&    219.8&      M&    10.14&     9.41&     8.97&     0.03&     0.04&     0.04&     0.76&     0.85&     0.83&      Y &    O-rich, member\\
     N6356& V4&    208.5&      M&    10.29&     9.49&     9.08&     0.03&     0.04&     0.04&     0.74&     0.79&     0.77&      Y &    O-rich, member\\
     N6388& V1&    184.4&      M&     9.91&     8.99&     8.64&     0.03&     0.03&     0.04&     0.66&     0.75&     0.73&      Y &    O-rich, member\\
     N6388& V2&    225.4&      M&     9.64&     8.73&     8.33&     0.03&     0.03&     0.03&     0.69&     0.70&     0.68&      Y &    O-rich, member\\
     N6388& V3&    157.6&      M&    10.33&     9.27&     9.00&     0.03&     0.03&     0.03&     0.44&     0.39&     0.38&      Y &    O-rich, member\\
     N6388& V4&    250.0&      M&     9.60&     8.82&     8.41&     0.04&     0.04&     0.05&     1.00&     1.12&     1.09&      Y &    O-rich, member\\
     N6441& V1&    201.0&      M&    10.29&     9.24&     8.90&     0.03&     0.03&     0.03&     0.47&     0.49&     0.48&      Y &    O-rich, member\\
     N6441& V2&    145.4&      M&    10.53&     9.59&     9.26&     0.03&     0.03&     0.03&     0.75&     0.65&     0.62&      Y &    O-rich, member\\
     N6441& V9&    142.8&   M/SR&    10.65&     9.61&     9.31&     0.03&     0.03&     0.03&     0.63&     0.56&     0.55&      Y &    O-rich, member\\
    N6441& V25&    210.0&     SR&    11.63&    10.07&     9.33&     0.03&     0.03&     0.03&     0.60&     0.57&     0.55&      N &    C-rich, CMD outlier, beyond $5r_h$\\
   N6441& V105&    110.6&   M/SR&    10.98&     9.88&     9.68&     0.03&     0.02&     0.03&     0.43&     0.29&     0.28&      Y &    O-rich, member\\
   N6441& V131&    128.4&      M&    10.93&     9.90&     9.64&     0.02&     0.02&     0.03&     0.42&     0.29&     0.29&      Y &    O-rich, member\\
   N6441& V135&    163.4&   M/SR&    10.65&     9.58&     9.30&     0.03&     0.03&     0.03&     0.71&     0.66&     0.64&      Y &    O-rich, member\\
   N6441& V139&    253.9&   M/SR&     9.90&     9.05&     8.60&     0.03&     0.03&     0.03&     0.78&     0.68&     0.66&      Y &    O-rich, member\\
    N6553& V4&    265.0&      M&     8.11&     7.10&     6.60&     0.04&     0.04&     0.04&     0.88&     0.92&     0.89&      Y   &   O-rich, member\\
    N6553& V5&    119.2&     SR&     8.06&     7.12&     6.63&     0.03&     0.03&     0.03&     0.58&     0.35&     0.34&      N & Possible non-LPV\\
    N6637& V1&    119.2&   M/SR&     9.68&     8.79&     8.55&     0.02&     0.02&     0.03&     0.22&     0.21&     0.21&      Y &    O-rich, small-amplitude variable\\
     N6637& V3&    254.0&     SR&     9.61&     8.80&     8.59&     0.02&     0.02&     0.03&     0.16&     0.17&     0.16&      N &    Field variable, CMD outlier\\
     N6637& V4&    198.0&      M&     8.98&     8.29&     7.93&     0.03&     0.03&     0.04&     0.74&     0.73&     0.72&      Y &    O-rich, member\\
     N6637& V5&    195.3&      M&     8.90&     8.14&     7.79&     0.03&     0.03&     0.04&     0.80&     0.78&     0.76&      Y &    O-rich, member\\
     N6637& V6&    141.4&     SR&     9.44&     8.51&     8.25&     0.02&     0.02&     0.03&     0.17&     0.16&     0.16&      N &    small-amplitude variable\\
     N6712& V2&    108.2&     SR&     9.06&     8.38&     8.11&     0.03&     0.03&     0.03&     0.71&     0.66&     0.65&      N &    CCD outlier\\
     N6712& V7&    191.2&      M&     8.72&     7.94&     7.53&     0.03&     0.03&     0.03&     0.71&     0.69&     0.67&      Y &   O-rich, member\\
     N6712& V8&    116.3&   M/SR&     9.37&     8.55&     8.27&     0.02&     0.02&     0.03&     0.25&     0.26&     0.23&      Y &    O-rich, member\\
     N6760& V3&    250.0&      M&     9.08&     8.16&     7.65&     0.04&     0.04&     0.04&     0.87&     0.85&     0.83&      Y &    O-rich, member\\
     N6760& V4&    227.4&      M&     9.36&     8.41&     7.89&     0.05&     0.05&     0.04&     1.20&     1.20&     1.06&      Y &    O-rich, member\\
     N6838& V1&    174.8&   M/SR&     7.55&     6.83&     6.52&     0.03&     0.03&     0.03&     0.46&     0.38&     0.37&      Y &    O-rich, member\\
      Pal6$^a$& V1&    560.0&      M&    14.01&    11.46&     8.70&     0.02&     0.07&     0.06&     --&     1.92&     1.59&      N &  Field variable, CMD/CCD outlier\\
     Ter12& V1&    463.7&      M&     8.74&     7.09&     6.18&     0.06&     0.05&     0.05&     1.54&     1.37&     1.17&      Y &    O-rich, $P > 400$ d\\
     Ter12& V2&    300.4&      M&    10.15&     8.83&     8.08&     0.04&     0.04&     0.04&     0.93&     0.88&     0.83&      N &    Field variable\\
      Ter5& V1&    255.7&      M&     8.96&     7.37&     6.56&     0.03&     0.03&     0.03&     0.60&     0.43&     0.42&      N &    Field variable, CMD outlier\\
      Ter5& V2&    216.8&      M&     9.78&     8.34&     7.61&     0.03&     0.04&     0.04&     0.78&     0.90&     0.88&      Y &    O-rich, member\\
      Ter5& V5&    452.6&      M&     9.90&     7.95&     6.78&     0.06&     0.05&     0.04&     1.55&     1.18&     0.96&      N &    Redder O-rich, CMD/CCD outlier\\
      Ter5& V6&    268.2&      M&    10.01&     8.38&     7.49&     0.03&     0.03&     0.04&     0.80&     0.77&     0.75&      Y &    O-rich, member\\
      Ter5& V7&    378.5&      M&     9.67&     8.03&     6.98&     0.04&     0.04&     0.04&     1.14&     0.87&     0.85&      Y & O-rich\\
      Ter5& V8&    261.3&      M&     9.79&     8.28&     7.44&     0.03&     0.04&     0.04&     0.81&     0.83&     0.82&      Y & O-rich, member\\
      Ter5& V9&    472.8&      M&    11.30&     9.47&     8.36&     0.05&     0.04&     0.04&     1.19&     0.85&     0.83&      N &   C-rich, CMD/CCD outlier\\
     Ter5& V10&    331.7&   M/SR&     9.35&     7.98&     7.17&     0.03&     0.03&     0.03&     0.70&     0.52&     0.50&      N &    Field variable\\
\enddata
\tablecomments{The cluster names are truncated throughout the text (see Table~\ref{tbl:ggcs} for full names). The `Flag' Yes/No refers to LPVs used in the initial PLR analysis. The remarks column provides reasons for exclusion of LPVs from the PLR fits. $^a$The $J$ band magnitude for Pal6 V1 is taken from 2MASS \citep{skrutskie2006}.}
\end{deluxetable}

To complement globular cluster variables, we also collected NIR time-series photometry for Milky Way field LPVs provided in \cite{whitelock2000} and \cite{whitelock2008}. Similar to cluster variables, these LPVs were cross-matched with the Gaia DR3 catalog to obtain their periods, amplitude, and C-star flag along with their astrometric information. We selected variables that have literature classification as Miras and a renormalized unit weight error (ruwe) less than 1.4, and have more than 10 photometric measurements in their NIR light curves. For this sample of 40 field Miras, we adopted homogeneous periods, photometry, and parallaxes from the Gaia catalogs. The NIR photometry for these Miras is in SAAO photometric system and is not transformed to 2MASS system. The field Mira sample will only be used for a comparative analysis and is not included for the Mira based distance scale in this work.

\subsection{Period and mean-magnitude determinations}\label{sec:subdata1}

We used multiband periodogram for sparsely sampled light curves from \citet{saha2017} to determine pulsation periods of LPVs. The period search was carried out between 100 and 1000 days for the IRSF light curves. The median difference between the estimated and Gaia periods is $0.2\%$, and the periods differ by more than $10\%$ only for N6388 V3 and N6441 V25. In both these cases, our period determinations agree well with those available on the updated \cite{clement2001} catalog. Despite being included in the Gaia LPV catalog, we did not find any significant variability or periodicity for N6352 V4, N6637 V3, and N6637 V6 in the IRSF sample. Among these, N6352 V4 and N6637 V6 have a small variability amplitude ($<0.2$~mag) in the $G$-band as well. The best period among Gaia and literature periods was selected for these three stars by visually inspecting their phased light curves. The period range of our sample is from 106 to 560 d, but the majority of LPVs have periods smaller than $\sim300$ d. 

The IRSF light curves of 55 LPVs are shown in Appendix \ref{sec:supp_fig} (Figures~\ref{fig:lpv_lc}--\ref{fig:lpv_lc2}). They are phased using their long-term periods. LPVs are not strictly periodic and our goal is to obtain their true mean magnitudes corresponding to the long-term period. Therefore, we fitted a sinusoidal template to the phased light curve in each filter assuming negligible phase lag between three NIR filters. The intensity-averaged mean magnitudes and peak-to-peak amplitudes were obtained from the best-fitted template light curves in $JHK_s$ filters.  Table~\ref{tbl:lpv_data} lists the pulsation properties of all 55 IRSF LPVs in our sample. The photometric properties of 40 field Miras in the Milky Way are also provided in Appendix~\ref{sec:supp_tbl} (Table~\ref{tbl:lpv_fielddata}). 

\subsection{Distances, reddening, and metallicity}\label{sec:subdata2}

\begin{deluxetable}{ccccc}
\tablecaption{Distance, metallicity, and reddening of LPV host globular clusters. \label{tbl:ggcs}}
\tabletypesize{\footnotesize}
\tablewidth{0pt}
\tablehead{
{Cluster Name} & {$d$ (kpc)} & {[Fe/H]} & {$E(B-V)$} & {$N_\textrm{lpv}/N_\textrm{plr}$}}
\startdata
   IC 1276&$    4.55\pm0.25   $ & $    -0.53\pm0.05    $&     1.08      &2/2\\
  Lynga 7&$    7.90\pm0.16   $ & $    -0.56\pm0.07    $&     0.73       &1/0\\
    NGC 104&$    4.52\pm0.03   $ & $    -0.71\pm0.01    $&     0.04     &5/5\\
    NGC 362&$    8.83\pm0.10   $ & $    -1.11\pm0.02    $&     0.05     &2/2\\
   NGC 5139&$    5.43\pm0.05   $ & $    -1.70\pm0.26    $&     0.12     &2/1\\
   NGC 5927&$    8.27\pm0.11   $ & $    -0.32\pm0.02    $&     0.45     &2/2\\
   NGC 6352&$    5.54\pm0.07   $ & $    -0.54\pm0.03    $&     0.22     &2/1\\
   NGC 6356&$   15.66\pm0.92   $ & $    -0.51\pm0.03    $&     0.28     &3/3\\
   NGC 6388&$   11.17\pm0.16   $ & $    -0.51\pm0.03    $&     0.37     &4/4\\
   NGC 6441&$   12.73\pm0.16   $ & $    -0.47\pm0.03    $&     0.47     &8/7\\
   NGC 6553&$    5.33\pm0.13   $ & $    -0.12\pm0.01    $&     0.63     &2/1\\
   NGC 6637&$    8.90\pm0.10   $ & $    -0.48\pm0.02    $&     0.18     &5/3\\
   NGC 6712&$    7.38\pm0.24   $ & $    -0.97\pm0.05    $&     0.45     &3/2\\
   NGC 6760&$    8.41\pm0.43   $ & $    -0.41\pm0.23    $&     0.77     &2/2\\
   NGC 6838&$    4.00\pm0.05   $ & $    -0.63\pm0.02    $&     0.25     &1/1\\
    Pal 6& $    7.05\pm0.45   $ & $    -0.85\pm0.02    $&     1.46      &1/0\\
   Terzan 12&$    5.17\pm0.38   $ & $    -0.47\pm0.16    $&     2.06        &2/1\\
    Terzan 5&$    6.62\pm0.15   $ & $    -0.40\pm0.02    $&     2.28        &8/4\\
\enddata
\tablecomments{The distances are from \citet{baumgardt2021}, metallicities are from \citet{dias2016}, and reddening values are taken from \citet{harris2010}. $N_\textrm{lpv}$ represents total number of IRSF LPVs in the cluster and $N_\textrm{plr}$ is the number included in the PLR analysis.}
\end{deluxetable}

The LPV sample was crossmatched with the globular cluster distances catalog provided by \citet{baumgardt2021}. There are 40 LPVs in 15 clusters that are within 10 kpc distance, and the median uncertainty on distance for the entire sample is $1.4\%$.
The mean parallaxes of globular clusters were also obtained from \cite{vasiliev2021} which were computed taking into account spatially correlated systematic errors after incorporating parallax corrections by \cite{lindegren2021}. Since several clusters in our sample are in the bulge, where crowding and extinction uncertainties dominate, the average uncertainty on the parallaxes is $10\%$. Furthermore, \cite{vasiliev2021} also suggested that parallax uncertainties are underestimated by $10-20\%$ and the parallaxes are overestimated by $\sim 0.01\pm 0.003$ mas for $G>13$ mag stars after applying \cite{lindegren2021} correction. The parallax overcorrection has also been noted for several other types of pulsating stars like RR Lyrae and Cepheids \citep{bhardwaj2021a, riess2021}. We also correct parallaxes for field Miras for the parallax zero-point correction from \cite{lindegren2021}. 

For the globular clusters, we obtained mean color-excess $E(B-V)$ from \citet{harris2010}. For the field Miras, the reddening was taken from the maps of \cite{schlegel1998} using the \texttt{dustmaps}\footnote{\url{https://dustmaps.readthedocs.io/en/latest/index.html}} package with updated calibration from \cite{schlafly2011}. The extinction correction was applied using the reddening law of \cite{fitzpatrick1999}, assuming an $R_V=3.1$. The total-to-selective absorption ratios: $R_{J/H/K_s} = 0.81/0.51/0.35$ \citep[e.g. in][]{bhardwaj2024} were estimated using the \texttt{dust$\_$extinction}\footnote{\url{https://pypi.org/project/dust-extinction/}} Python package. We also obtained mean metallicities of globular clusters from the catalog of \citet{dias2016}. The distances, metallicities, and reddening of the LPV host clusters are provided in Table~\ref{tbl:ggcs}.

\section{Luminosity calibration of Miras}
\label{sec:lpv_plr}

Among LPVs, O-rich Mira variables follow a tight PLR, but a significant contamination comes from C-rich Miras at longer periods \citep{yuan2017b, huang2018}. 
In the long-period end, O-rich Miras also display a broken or quadratic PLR due to additional luminosity arising from the hot-bottom burning in AGB stars \citep{whitelock2003}. Therefore, it is important to identify short-period ($P<400$ d) O-rich LPVs to derive their accurate and precise PLRs in the NIR bands. 

\subsection{Selection of O-rich Miras and cluster membership}

\begin{figure*}[t!]
\centering
\includegraphics[width=0.98\textwidth]{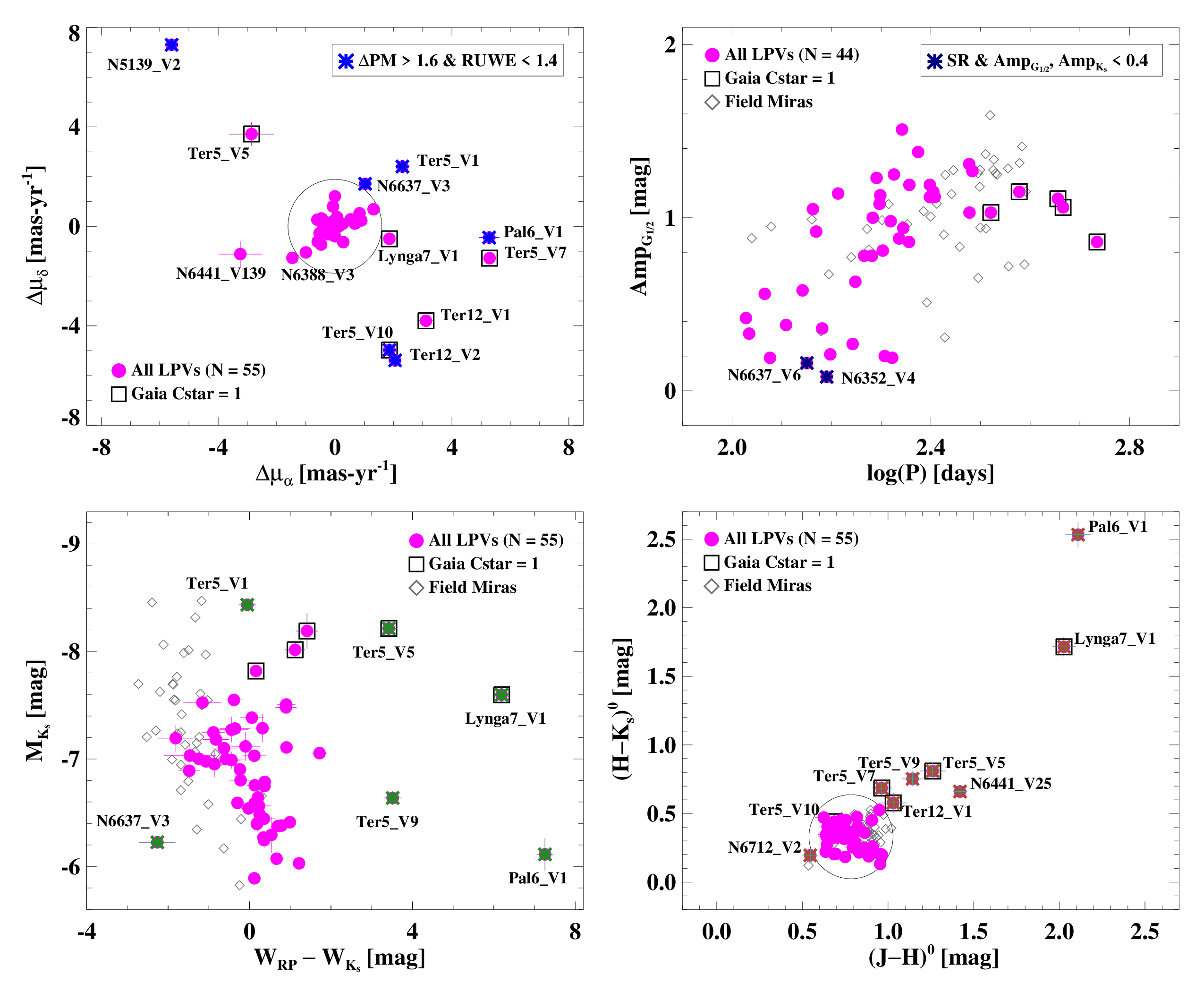}
\caption{{\it Top left:} the differences in the proper motions between LPVs and their host globular clusters along right ascension and declination axes. LPVs with good astrometric solution (ruwe $< 1.4$) that are likely field stars are marked in blue asterisks. {\it Top right:} Half of the $G$-band variability amplitude for cluster and field LPVs taken from \citet{lebzelter2023}. LPVs with SR classification or low optical-NIR amplitudes are marked in colored asterisk symbols. {\it Bottom left:} The Wesenheit color and magnitude diagram for LPVs based on mean magnitudes and colors as defined in \citet{lebzelter2018}. {\it Bottom right:} Extinction corrected color-color diagram for LPVs is shown. The outliers (C-rich candidate/reddened O-rich variables or potential non-cluster member or low amplitude/SR variables) are shown with colored asterisk symbols.}
\label{fig:pm_cmd}
\end{figure*}

We use Gaia astrometry, mean optical and NIR magnitudes, amplitudes, and colors to confirm cluster membership of stars and select O-rich Miras. Top left panel of Fig.~\ref{fig:pm_cmd} displays the difference between proper motions of LPVs and mean proper motion of their host cluster. The majority of stars have differences between their individual and cluster proper motions close to zero along both the right ascension and declination axes. We estimated a scatter of 0.533 mas--yr$^{-1}$ around the mean difference values and considered LPVs with proper motion residuals beyond a radius of $3\sigma\sim1.6$~mas--yr$^{-1}$ as potential field stars.  Since several LPVs are in the bulge globular clusters where crowding is severe and extinction is relatively high, their proper motion accuracy may not be reliable for membership analysis. There are 14 LPVs with ruwe $>$ 1.4, and these were not considered in membership analysis. Among those with ruwe $< 1.4$, we identified 6 LPVs (in blue asterisks -- Pal6 V1, N5139 V2, N6637 V3, Ter5 V1, Ter5 V10, and Ter12 V2) as likely non-members of their host cluster. Furthermore, N5139 V2 is also located beyond $5r_h$ of their cluster in the spatial distribution.

Top right panel of Fig.~\ref{fig:pm_cmd} displays half of the peak-to-peak amplitude of best-fit model to the Gaia $G$-band light curves of the LPVs \citep{lebzelter2023}. Considering photometric uncertainties on individual measurements ($\sim0.05$ mag) in the IRSF light curves, we do not adopt a strict threshold of $0.4$ mag only in NIR amplitudes for Miras. The cluster LPVs have a wide range of metallicities, which may impact their variability amplitude to separate Mira variables. There are also a few field Miras with Amp$_{G_{1/2}} < 0.4$ mag. In contrary, N6637 V3 exhibits Gaia variability amplitude exceeding 1 mag, but no significant variability is seen in NIR light curves. 
However, two LPVs have low optical and infrared amplitudes ($<0.2$ mag, N6352 V4 and N6637 V6) and are classified as SR variables. These LPVs could be small amplitude stars and were excluded from the selection of O-rich Mira variables.

The bottom panels of Fig.~\ref{fig:pm_cmd} utilize color-magnitude and color-color diagrams to identify possible C-rich Miras. Following \citet{lebzelter2018}, we used difference of optical and NIR Wesenheit magnitudes, $W_{RP} = G_{RP} - 1.3 (G_{BP} - G_{RP})$ and $W_{K_s}= K_s - 0.686(J - K_s)$, as a proxy for color together with their absolute $K_s$ magnitude to separate O-rich and C-rich AGB stars. Four LPVs (Lynga7 V1, Pal6 V1, Ter5 V5, and Ter5 V9) are distinctly redder, which is also evident in the color-color diagram on the bottom right panel. N6637 V3 and Ter5 V1, the field variables, are also outliers in the color-magnitude plane. In the color-color diagram, N6441 V25, and N6712 V2 are also beyond $3\sigma$ circular radius in the colors along both axes, where $\sigma=0.09$ mag is the scatter around the median color values. Several bulge globular clusters, in particular Terzan 5 \citep{massari2012}, exhibit significant differential reddening, and a classification of individual stars based on color-magnitude planes may be sensitive to our assumption of average reddening in the cluster. 

In summary, we confirm Lynga7 V1 as a C-rich variable \citep{feast2013}, Ter5 V5 is a known O-rich Mira \citep{matsunaga2005}, but it is either significantly reddened or is a field star. Pal6 V1, N5139 V2, N6637 V3, Ter5 V1, Ter5 V10, and Ter12 V2 are most likely field LPVs based on their proper motions. Ter5 V9 and N6441 V25 are possibly C-rich Mira variables and N6352 V4 and N6637 V6 may be small amplitude variables. These 12 LPVs are flagged as potential outliers for the O-rich Mira PLR analysis in the following section. 

\subsection{Mira Period-Luminosity relations}

The extinction corrected mean magnitudes of LPVs together with their cluster distances from \citet{baumgardt2021} were used to fit PLRs in the following linear form:

\begin{equation}
M_\lambda = a_\lambda + b_\lambda(\log P - 2.3),
\end{equation}

\noindent where, $M_\lambda$ is the absolute magnitude in $JHK_s$ bands, and $a_\lambda$ and $b_\lambda$ represent the zero-point and slope, respectively, at a given wavelength. We obtained a preliminary linear regression fit to our sample of 55 LPVs after iteratively removing $3\sigma$ outliers. We found 8 LPVs (Lynga7 V1, Pal6 V1, N6441 V25, N6553 V5, N6637 V3, Ter5 V1, Ter5 V9, Ter12 V2) exhibiting residuals larger than 0.5 mag in all three $JHK_s$ filters. Seven of these eight LPVs have been identified as outliers in the previous subsection, confirming their exclusion as non-cluster members or potential C-rich stars or highly reddened O-rich Miras. The residuals of N6553 V5 exceed 0.8 mag in NIR PLRs, suggesting its period of 119.2, estimated from sparsely sampled light curves, may not be correct. This variable was not classified as LPV in the Gaia and does not have any period determination in the \citet{clement2001} catalog. Therefore, N6553 V5 may not be an LPV and is excluded from the subsequent analysis.  

\begin{figure}
\centering
\includegraphics[width=0.49\textwidth]{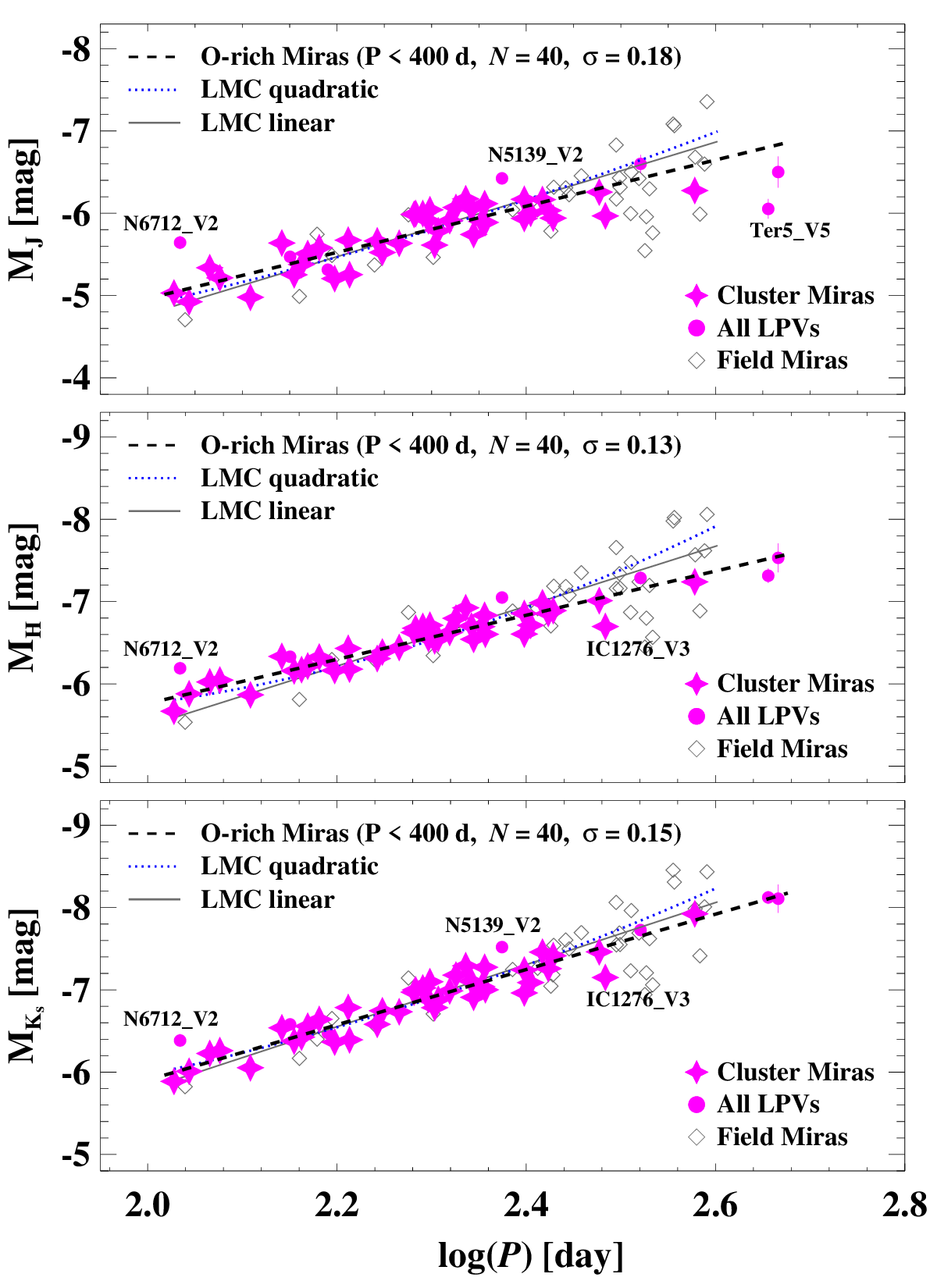}
	\caption{$JHK_s$ period-luminosity relations for 47 LPVs in globular clusters after excluding significant outliers in Fig.~\ref{fig:pm_cmd} (see text for details). The black dashed line represents the best-fitting linear regression to 40 O-rich Mira candidate variables (star symbols) with periods smaller than 400 days. The relation is extrapolated to the entire period range. Smaller circles represent LPVs that are excluded from the PLR fits. NIR PLRs for LMC Miras are also shown in blue dotted (quadratic relation) and solid grey line (linear relation) from \citet{yuan2017b}.} 
\label{fig:lpv_plr}
\end{figure}

After excluding eight LPVs with the largest residuals, the PLRs for 47 variables are shown in Fig.~\ref{fig:lpv_plr} in $JHK_s$ bands. Among five remaining outliers identified in the previous subsection, N5139 V2 and Ter5 V5 are marked, while Ter5 V10 does not show residuals beyond $2\sigma$ despite being a field variable. Similarly, N6352 V4 and N6637 V6 are also fully consistent with the PLR fit but are excluded from the final analysis due to their small or null amplitude variations ($\sim0.1$~mag). One candidate SR variable, N6712 V2 is a $2\sigma$ outlier in all three $JHK_s$ bands and is also bluer than the typical colors of O-rich stars in our sample (see Fig.~\ref{fig:pm_cmd}). All of these six LPVs (N5139 V2, Ter5 V5, Ter5 V10, N6352 V4, N6637 V6, and N6712 V2) were also excluded from the final sample of 41 O-rich cluster member Mira candidates. This sample contains 30 Miras, 11 M/SR variables, among which N6637 V1 has the lowest NIR amplitudes ($\sim 0.2$ mag).

\begin{deluxetable}{cccccc}
\tablecaption{Near-infrared PL and PW relations of Miras in globular clusters in our homogeneous IRSF sample. \label{tbl:plr_mira}}
\tablewidth{0pt}
\tablehead{
{~~Band~~} & {~~$\alpha_\lambda$~~} & {~~$\beta_\lambda$~~} & {~~$\sigma$~~}& {~~$N_i$~~} & {~~$N_f$}~~}
\startdata
    $J$ &$     -5.81\pm0.03      $ & $     -2.81\pm0.24       $&       0.18 &   41&  40\\
    $H$ &$     -6.57\pm0.02      $ & $     -2.68\pm0.16       $&       0.13 &   41&  40\\
  $K_s$ &$     -6.91\pm0.02      $ & $     -3.36\pm0.19       $&       0.15 &   41&  40\\
          $W_{J,H}$ &$     -7.85\pm0.03      $ & $     -2.57\pm0.23       $&       0.19 &   41&  40\\
      $W_{J,K_s}$ &$     -7.75\pm0.03      $ & $     -3.86\pm0.20       $&       0.17 &   41&  40\\
      $W_{H,K_s}$ &$     -7.67\pm0.04      $ & $     -4.93\pm0.28       $&       0.23 &   41&  40\\
\enddata
\tablecomments{The zero-point ($a$), slope ($b$), dispersion ($\sigma$) and the number of stars ($N_{i/f}$) in the initial/final PL fits are tabulated. Only excluded Mira variable (Ter12 V1) from the PLR fit has period longer than 400 d.}
\end{deluxetable}

The final PLR fit for 40 O-rich Mira candidates was obtained excluding the only remaining LPV with $P > 400$ days (Ter12 V1). The best-fitted linear regression (dashed lines in Fig.~\ref{fig:lpv_plr}) was extrapolated over the entire period range, and compared with the linear and quadratic relations for the O-rich Miras in the LMC \citep{yuan2017b}. The cluster Miras exhibit a shallower slope in $JHK_s$ filters, but agree well with LMC relations in the period range from 150 to 350 days. While the difference in slopes between cluster and LMC Mira PLRs is statistically significant in $JH$ bands, the zero-points differ by $< 1\sigma$ in all cases. The difference in slopes is likely due to small number of LPVs at the longer period end as there are only 3 stars with periods between 300 and 400 d in the cluster sample. 

The results of the best-fit PLRs based on IRSF sample of 40 Mira candidates are tabulated in Table~\ref{tbl:plr_mira}. The PLRs in $JHK_s$ bands exhibit scatter comparable to Miras in the LMC. 
Table~\ref{tbl:plr_mira} lists statistical errors only that were determined using bootstrapping procedure with $10^4$ random realizations of PLR fits. We adopt the average uncertainty of 0.029 mag as systematic errors in the globular cluster distances which corresponds to $1.4\%$ at the mean distance of 8.4kpc for Miras in our final IRSF sample. 

 To investigate the impact of reddening corrections, we also derive PW relations in NIR filters, defined as: $W_{J,H} = H - 1.67 (J - H)$, $W_{H,K_s} = K_s - 2.19 (H - K_s)$, and $W_{J,K_s} = K_s - 0.75(J - K_s)$, based on the \cite{fitzpatrick1999} extinction law (see Section~\ref{sec:subdata2}). The slope and zero-points of the period-Wesenheit relations (PWRs) are consistent between the LMC and cluster Miras, except for $W_{J,H}$, in which the slopes are statistically different. Note that $W_{J,H}$ PWRs exhibit comparably large scatter in both the LMC and globular cluster. These Wesenheit color coefficients are different from those adopted by \citet{yuan2017b}, and are not necessarily appropriate as the interstellar extinction law may differ from the circumstellar dust extinction in Mira variables. 

We further considered different variants of the PLR analysis and found that the coefficients of these relations do not change significantly. These variants include -- 1) amplitude cuts (Amp$_{K_s} > 0.3/0.4$~mag); 2) use of different reddening law from \citet{fitzpatrick1999}; 3) a different  assumption of $R_V=3.23$; 4) restricting sample to low-reddened clusters ($E(B-V) < 1$) mag; and 5) using only Miras with `M' classification (29 stars excluding M/SR variables).

Since our sample of cluster Miras covers a wide metallicity range ($-1.7<$ [Fe/H] $<-0.1$), we investigated residual variations and found no significant trend as a function of metallicity. To quantify possible metallicity effects, we also solved for an additional metallicity term in the PLRs and PWRs for Mira variables in clusters and found a null or negligible dependence ($\lesssim 0.002\pm0.002$ mag/dex) on iron abundance in $JHK_s$ bands. We also separated the O-rich Mira sample into the metal-rich ([Fe/H]$>-0.5$~dex, 21 stars) and metal-poor ([Fe/H]$<-0.5$~dex) samples and derived their separate PLRs. While the slopes are similar within $1\sigma$ for the two subsamples, the zero-points are statistically consistent within $2\sigma$ in all three $JHK_s$ bands. We conclude there is no significant metallicity effect on the PLRs for cluster Mira variables. A similar null or small effect of metallicity has also been noted in previous studies on Miras in globular clusters and LMC, and AGB stars in nearby metal-poor galaxies \citep{rejkuba2004, whitelock2008, bhardwaj2019, goldman2019}. However, the majority of LPVs in our sample belong to clusters with a narrow range of metallicity, $-0.7 < $[Fe/H]$ < -0.3$, which can limit a proper quantification of metallicity dependence of PLRs for cluster Mira variables.

\subsection{Astrometry-based luminosity fits}

The absolute calibration of Mira PLRs listed in Table~\ref{tbl:plr_mira} is based on recommended distances to the clusters by \citet{baumgardt2021}. However, homogeneous Gaia DR3 parallax-based distances are not available for the entire sample of clusters in this catalog. Therefore, we also use mean parallaxes for globular clusters to calibrate NIR PLRs for Mira variables \citep{vasiliev2021}. As mentioned in  Section~\ref{sec:subdata2}, the average uncertainty on the mean parallaxes of clusters in our sample is $10\%$. Therefore, the calibration is obtained using astrometry-based luminosity \citep[ABL,][]{feast1997,arenou1999} defined as:

\begin{eqnarray}
 {\textrm{ABL}} ~&=&~ \overline{\omega}_\textrm{(mas)}10^{0.2m_\lambda-2} \nonumber \\
~&=&~ 10^{0.2(\alpha_{\lambda} + \beta_{\lambda} (\log P - 2.3))},
    \label{eq:plr_abl}
\end{eqnarray}

\noindent where $m_\lambda$ is the extinction corrected apparent mean magnitude in $JHK_s$. The results of the ABL fits are shown in Table~\ref{tbl:plr_abl} after iteratively removing the $3\sigma$ outliers. The slopes and zero-points obtained using the ABL fits are within $2\sigma$ of those listed in Table~\ref{tbl:plr_mira} in all cases, showing consistency with the PLR calibrations based on globular cluster distances. Nevertheless, the zero-points based on this approach  systematically differ by 0.05/0.06/0.04 mag in $J/H/K_s$, hinting an overcorrection in cluster parallaxes as suggested by \citet{vasiliev2021}. 

We also used the field Mira sample to derive PLRs in NIR bands based on ABL method. The results of the best-fit listed in Table~\ref{tbl:plr_abl} show a better agreement ($<1\sigma$) with zero-points of PLRs obtained using globular cluster distances. The field Mira distances are distributed between 0.3 and 3.8 kpc, and their median uncertainties ($6\%$) are smaller than errors on mean cluster parallaxes. This results in a better agreement in zero-points of PLRs for field Miras and their cluster counterparts for which the calibration is based on \cite{baumgardt2021} distances. 
However, if parallaxes are transformed to distance space, the scatter in the field Mira PLRs is nearly two times larger than the values quoted in Table~\ref{tbl:plr_mira} for cluster variables. This suggests that the parallaxes for Mira variables even in solar neighborhood may not be very accurate and reliable for the precise calibration of PLRs using field stars.

\begin{deluxetable}{cccccc}
\tablecaption{Astrometry-based luminosity fits for PLRs of Mira variable stars. \label{tbl:plr_abl}}
\tablewidth{0pt}
\tablehead{
{~~Band~~} & {~~$\alpha_\lambda$~~} & {~~$\beta_\lambda$~~} & {~~$\sigma$~~}& {~~$N_i$~~} & {~~$N_f$}~~}
\startdata
\hline
\multicolumn{6}{c}{Globular cluster Miras with $P<400$ d}\\
\hline
              $J$ &$    -5.76\pm0.03    $&$    -2.40\pm0.25    $&    0.010&  40& 39\\
              $H$ &$    -6.51\pm0.03    $&$    -2.53\pm0.25    $&    0.004&  40& 37\\
            $K_s$ &$    -6.87\pm0.03    $&$    -3.10\pm0.25    $&    0.004&  40& 37\\
\hline
\multicolumn{6}{c}{Milky Way field Miras with $P<400$ d}\\
\hline
              $J$ &$    -5.79\pm0.02    $&$    -3.06\pm0.11    $&    0.010&  40& 40\\
              $H$ &$    -6.58\pm0.02    $&$    -3.42\pm0.11    $&    0.006&  40& 40\\
            $K_s$ &$    -6.89\pm0.02    $&$    -3.79\pm0.11    $&    0.005&  40& 40\\
\enddata
\tablecomments{The zero-point ($a$), slope ($b$), dispersion ($\sigma$) and the number of stars ($N_f$) in the final PL fits are tabulated. The initial sample ($N_i$) consists of 40 O-rich Miras excluding Ter12 V1. The MW field Mira zero-points are in SAAO photometric system.}
\end{deluxetable}

\subsection{Distance to the LMC}

To validate the accuracy of the Mira PLRs listed in Table~\ref{tbl:plr_mira} for the distance scale, we use these calibrations to obtain a distance to the LMC. The photometric mean magnitudes of Miras in the LMC were taken from \citet{yuan2017b}. This sample was restricted to 163 O-rich Miras with periods smaller than 400 d. The period distribution of O-rich Miras in the LMC adequately covers the range between 100 and 400 d, unlike the paucity of cluster counterparts between 300 and 400 d. Therefore, we adopt the slopes of LMC Mira PLRs in $JHK_s$: $b_J=-3.48\pm0.09$, $b_H=-3.64\pm0.09$, and $b_{K_s}=-3.77\pm0.07$ \citep{yuan2017b} as a reference to determine apparent zero-point in the LMC and a calibrated zero-point in the globular clusters. 

We note that the LMC Mira sample was not corrected for interstellar reddening. Therefore, we estimate color-excess, $E(V-I)$, from the reddening maps of \cite{skowron2021}, and find a median value of $E(V-I)=0.102\pm0.066$~mag at the location of O-rich Miras. The reddening value was converted to $E(B-V)=0.08$ in \cite{schlegel1998} scale using transformations from \citet{skowron2021}. Mira magnitudes were corrected for extinction using the \cite{fitzpatrick1999} reddening law as discussed in Section~\ref{sec:subdata2}. The results of fitting the fixed slopes in $JHK_s$ are shown in Table~\ref{tbl:lmc_dis}, where the LMC distance agrees well with the most precise value of $18.477\pm0.026$ from late-type eclipsing binaries \citep{piet2019}. The uncertainty on the LMC distance modulus includes the systematics of 0.029 mag in calibrator cluster distances. The $J$-band distance is smaller by $\approx1\sigma$ than that derived from $H$ and $K_s$ band PLRs. The average true distance modulus of $18.45\pm0.04$~mag using $HK_s$ bands is in very good agreement with the geometric distance to the LMC \citep{piet2019}, thus establishing the use of Miras in globular clusters to constrain their accurate luminosity zero-points in the NIR bands. 

\begin{deluxetable}{cccc}
\tablecaption{LMC distance. \label{tbl:lmc_dis}}
\tablewidth{0pt}
\tablehead{
{Band} & {$a_\lambda$ (LMC)} & {$a_\lambda$ (GC)}& {$\mu$ (LMC)}}
\startdata
              $J$ &$    12.59\pm0.01    $&$    -5.82\pm0.03    $&$    18.41\pm0.04    $\\
              $H$ &$    11.86\pm0.01    $&$    -6.60\pm0.03    $&$    18.46\pm0.04    $\\
            $K_s$ &$    11.52\pm0.01    $&$    -6.92\pm0.03    $&$    18.44\pm0.04    $\\
\enddata
\tablecomments{LMC and GC zero-points are corrected for reddening.}
\end{deluxetable}

\section{Hubble Constant Determination}
\label{sec:h0}

The local determination of the Hubble constant is typically obtained using a three-step distance ladder where in the first step, primary calibrators like Cepheids and Miras are used to derive their absolute PLRs. 
Precise distances based on Cepheids or Miras in the second step provide a calibration of SNe Ia peak luminosity. This is used as reference in the third step, where SNe probe the expansion rate of the Universe, to determine the Hubble constant \citep{riess2022, freedman2024, huang2024}. The following subsections discuss these three different steps for the Mira based distance scale in this work. 

\subsection{Ground-based $H$ to {\it HST} F160W PLR}

\begin{figure}
\centering
\includegraphics[width=0.48\textwidth]{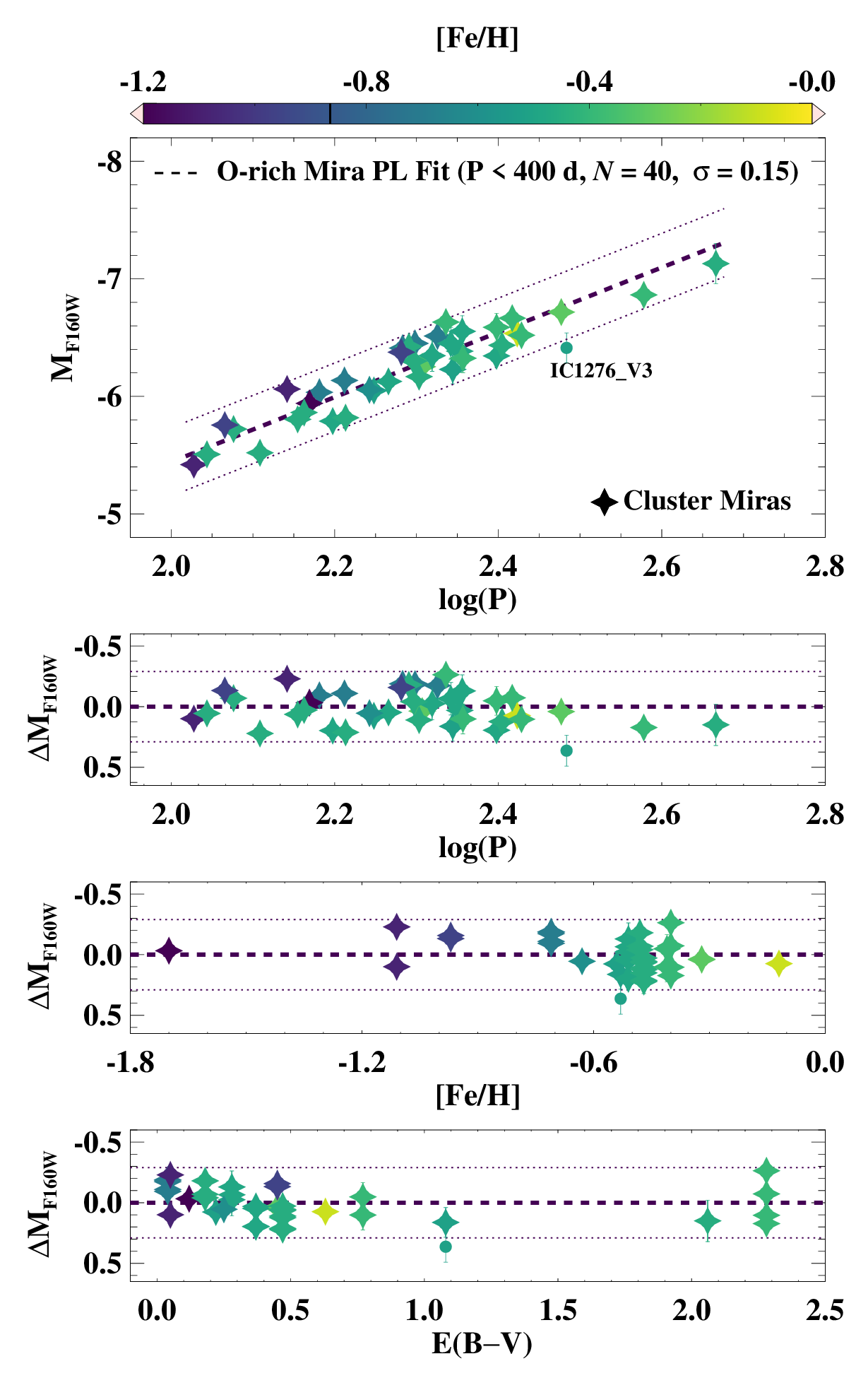}
	\caption{{\it Top:} absolute calibration of PLR in {\it HST} F160W filter for 41 O-rich cluster Miras. The dashed line represents the best fitting linear regression for 40 Miras with periods smaller than 400 d. The small circle shows a Mira with residuals larger than $\pm2\sigma$ dotted lines. {\it Bottom panels:} the residuals of Mira PLR fit are shown as a function of their periods, and metallicity and reddening of their host cluster.} 
\label{fig:hst_plr}
\end{figure}

To complement the observations of Miras in SNe Ia host galaxies in the {\it HST} F160W photometric system, \cite{huang2018} derived a color transformation: $F160W = H + 0.39(J - H)$. In the case of globular clusters, we used extinction corrected mean magnitudes in $J$ and $H$ bands transforming $H$ to F160W magnitudes. Using the same initial sample of 41 Miras as in Table~\ref{tbl:plr_mira}, a PLR in the F160W filter was derived:
\begin{equation}
M_{F160W} =  -6.27(0.02) - 2.76(0.18)(\log P - 2.3),\\
\end{equation}

\noindent with a scatter of 0.15 mag for stars with periods smaller than 400 d. Fig.~\ref{fig:hst_plr} displays this PLR in the {\it HST} F160W filter. We do not find any trends in the residuals of the best-fitting relation as a function of periods of Miras, and with metallicity or reddening of their host clusters. The globular clusters cover a wide metallicity range $-1.7 < \textrm{[Fe/H]} < -0.1$, thus these results further show that cluster Mira PLRs are not sensitive to metallicity effects as previously suggested for Mira/AGB stars in several studies at infrared wavelengths \citep{whitelock2008, bhardwaj2019, goldman2019}. 

\subsection{The global slope and zero-points of anchor PLRs}\label{sec:anchor_zp}

\begin{figure*}
\centering
\includegraphics[width=0.98\textwidth]{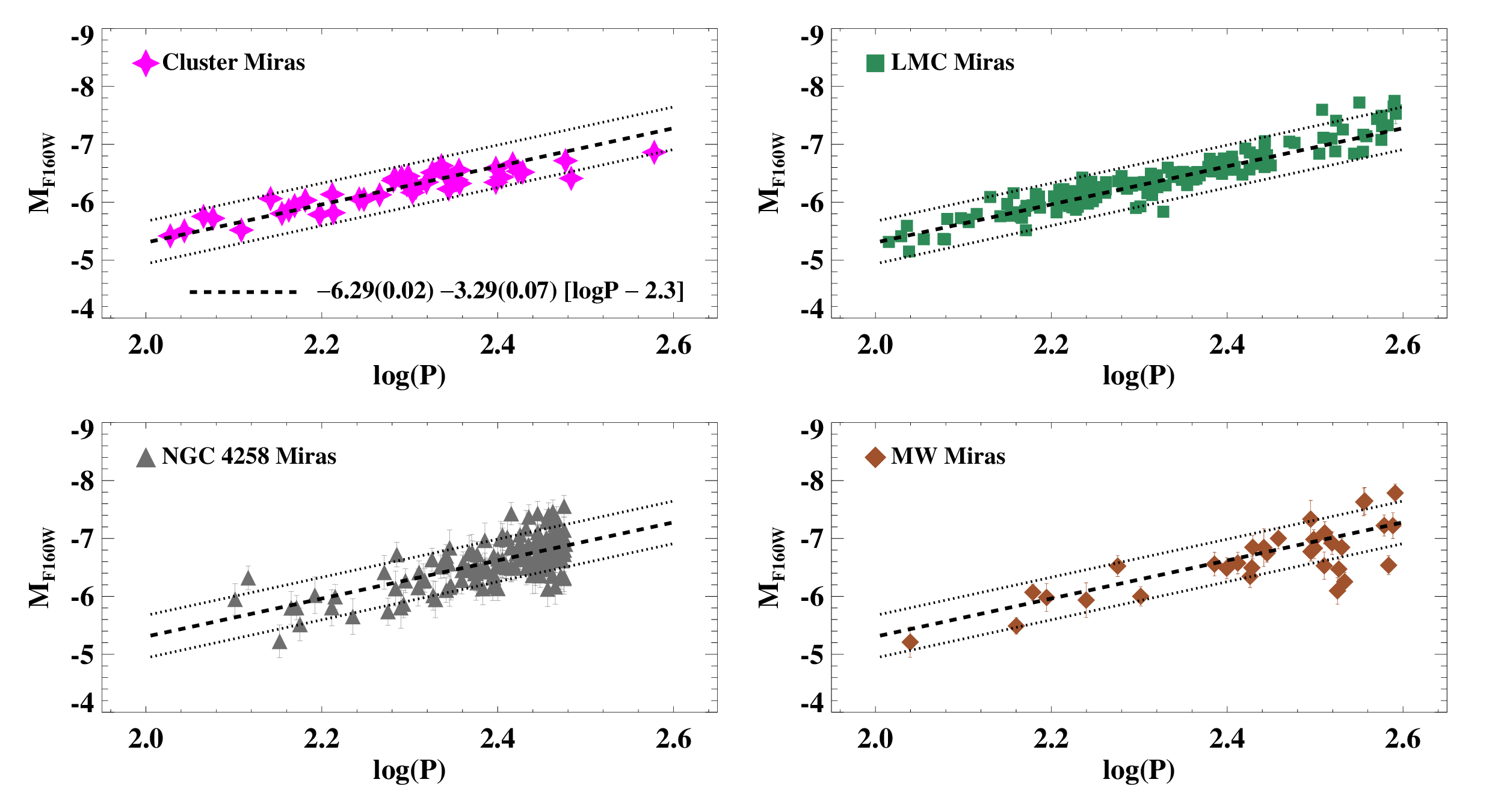}
	\caption{Mira PLRs in F160W filter in three anchors -- Globular clusters, LMC, and NGC 4258. The Milky Way field Miras are also shown in the bottom-right panel, but were not included in the global analysis. The global PLR shown in the top-left panel represents the best fitting linear regression to Miras with $P<400$ d in three anchor galaxies. The dashed lines represent the global slope fitted to all F160W PLRs and the dotted lines display $\pm2\sigma$ scatter.} 
\label{fig:all_plr}
\end{figure*}

The universality of the slope of Mira PLR in F160W filters is the basic assumption for the distance scale application. \citet{huang2018} used two different slopes of $-3.64$ mag/dex based on \citet{yuan2017b} LMC sample, and $-3.35$ mag/dex based on the transformed {\it HST} F160W filter. The slopes of the globular cluster and LMC PLRs in the {\it HST} F160W filter are significantly different, presumably due to different period distribution of Miras in these systems. With three independent anchors available, we can determine a global slope of Mira PLRs in calibrated F160W magnitudes. For the LMC, we adopt the $1.2\%$ precise geometric distance to the LMC \citep{piet2019}. For the NGC 4258, a geometric maser distance of $29.397\pm0.032$~mag \citep{pesce2020} was adopted that is precise to $1.5\%$. 

Fig.~\ref{fig:all_plr} displays PLRs for Miras in F160W filter in anchor galaxies. The Milky Way field Miras discussed in Section~\ref{sec:subdata2} are also shown, but were not included in the global fit as their parallaxes exhibit significantly larger mean uncertainty of $6\%$. The slope of the global fit was found to be $-3.29\pm0.07$~mag, in good agreement with the adopted slope in \cite{huang2024}. 

\begin{figure}
\centering
\includegraphics[width=0.49\textwidth]{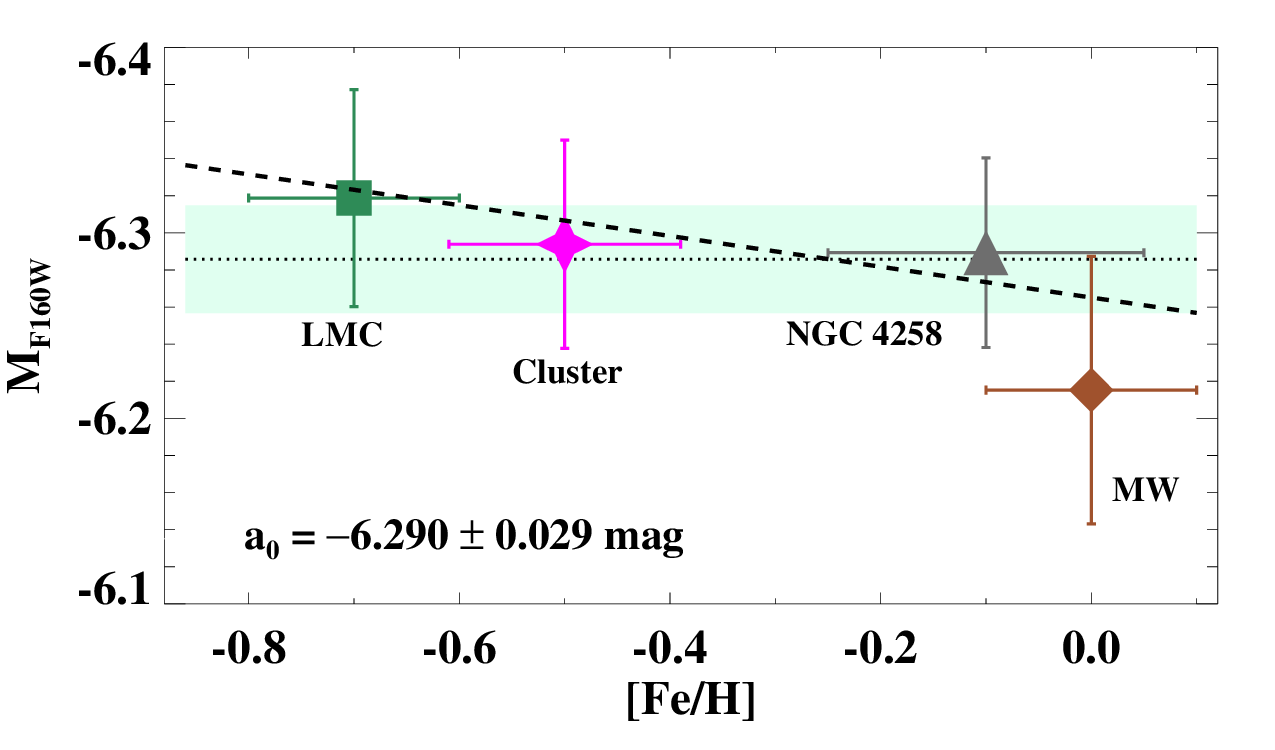}
	\caption{The zero-point variation among anchor PLRs for the fixed global slope as a function of metallicity. The dotted line shows a weighted variance mean of the Cluster, LMC, and NGC 4258 zero-points. The dashed line represents the two-parameter linear-fit to these three anchor solutions (eq.~\ref{eqn:fit}) with a metallicity slope of $0.09\pm0.04$ mag/dex. The shaded region represents $\pm1\sigma$ error on the absolute zero-point ($a_0$). The error in zero-points include anchor distance, metallicity systematics, and ground to HST transformation uncertainties (see text for details).} 
\label{fig:all_met}
\end{figure}

To account for the possible metallicity effects, we adopt a two-parameter solution simultaneously deriving a zero-point and a metallicity coefficient. For the cluster Miras, we derive a mean metallicity of $-0.50$ with a scatter of $0.11$~dex after excluding the most metal-poor cluster. For the LMC, a metallicity of $-0.7\pm0.1$~dex is taken based on age-metallicity relations for intermediate-age stellar populations in the field and cluster stars \citep{narloch2023}. \cite{feast1996} also suggested [Fe/H]$\sim-0.6$~dex for Miras in the LMC based on a comparison with cluster variables. Given the lack of metallicity measurements for stellar populations in NGC 4258, we adopt a value of $-0.10\pm0.15$~dex based on H\texttt{II} region emission line flux measurements from \citet{yuan2022}. This value is consistent with stellar metallicity of blue supergiants \citep{kudritzki2024} and has been used in the Cepheid-based distance ladder \citep{riess2022}. While the adopted metallicity is appropriate for younger stellar populations, considering no significant metallicity gradient in NGC 4258 \citep{yuan2022, kudritzki2024} and the lack of stellar metallicity measurements, this approximation is adopted with a relatively larger uncertainty of 0.15 dex. For the Milky Way (MW) field stars, we assume solar metallicity for Miras, but this was not included in the quantitative analysis. Using the aforementioned fixed slope, we estimated the following global solution:

\begin{eqnarray}
    M_{F160W} &=& -6.29(0.02) - 3.29(0.07)[\log P - 2.3]  \nonumber \\
    & &+ ~0.09(0.04)(\textrm{[Fe/H]}-\textrm{[Fe/H]}_\textrm{mean}),
\label{eqn:fit}
\end{eqnarray}

\noindent where the mean metallicity of the anchor sample is [Fe/H]$_\textrm{mean}=-0.42$~dex. We find a small positive metallicity coefficient such that metal-rich Miras are fainter.  However, this quantification of metallicity effect is based on our approximation of mean metallicities of Miras in the LMC, in globular clusters and NGC 4258. Fig.~\ref{fig:all_met} displays this linear fit to the zero-points of three anchor galaxies as a function of their adopted metallicities. We find a small variation in the zero-point of $<0.02$~mag with or without the metallicity term, further suggesting a small effect in the luminosity due to variations in iron abundance. 

Furthermore, no metallicity measurements of Miras or intermediate-age populations are available in SNe Ia host galaxies. \citet{huang2024} used an oxygen abundance gradient based-on H\texttt{II} regions to derive an approximate metallicity range for M101 Miras that falls between solar and LMC values. We assume that the SNe Ia host galaxies have metallicities similar to NGC 4258 with a possible difference of $\pm0.2$~dex, which may increase up to $\pm0.5$~dex with respect to LMC Miras. Given the approximations involved in determining the metallicity coefficient, we do not include this term in the subsequent analysis, but add a systematic uncertainty term accounting for the difference in mean metallicities between anchor and host galaxies.

Using the fixed slope, we find $a_{GGC}=-6.294\pm0.056$, $a_{LMC}=-6.319\pm0.058$, $a_{NGC 4258}=-6.289\pm0.051$~mag, and $a_{MW}=-6.242\pm0.071$. In addition to statistical uncertainties, the errors on the zero-points include uncertainties due to the anchor distance, $H$ to F160W transformation (0.02 mag for the LMC/Cluster/MW), and metallicity systematics due to aforementioned metallicity differences with respect to SNe Ia host galaxies (0.036/0.045/0.018/0.018 for the Cluster/LMC/NGC 4258/MW). A weighted-variance average of three independent anchors (MW is not included), $a_0 = -6.290\pm0.029$~mag, was adopted as the calibrated zero-point magnitude of a Mira with $\log P = 2.3$ in the first rung of the distance ladder. The absolute zero-point is in excellent agreement with the result of \citet{huang2024}, but comes from the baseline solution of three independent anchor galaxies. 

\subsection{Distances to Type Ia Supernovae host galaxies} \label{sec:sne_dis}

The second step of the distance ladder involves using primary calibration of stellar standard candles to determine distances to SNe Ia host galaxies and calibrate the peak fiducial luminosity of SNe Ia. For the Mira distance scale, there are two supernovae host galaxies observed with {\it HST} --- SN 2005df in NGC 1559 \citep{huang2020} and SN 2011fe in M101 \citep{huang2024}. We used a sample of 110 Miras in NGC 1559 and 211 Miras in M101 from the aforementioned studies along with their periods and calibrated F160W magnitudes including corrections due to crowding effects.
We applied corrections for the Milky Way foreground extinction as described in \citet{huang2024}.

Following \citet{huang2024}, the calibrated fiducial peak luminosity of SN Ia was obtained in the following two steps: we first determined the apparent zero-points of Mira PLRs in SNe Ia hosts by adopting the global slope of the PLRs in the anchor galaxies: 
\begin{equation}
m_{F160W} = a_{targets} + 3.29(\log P - 2.3). 
\end{equation}

We estimated the apparent zero-points of $a_{N1559}=25.098\pm0.035$ and $a_{M101}=22.780\pm0.023$. The errors include an uncertainty of 0.02 mag due to interstellar extinction in SNe host galaxies and the variation in the adopted slope (0.01 mag) for the anchor PLR. Using the absolute zero-point of Miras in Section~\ref{sec:anchor_zp}, we also obtained distances to SNe host galaxies $\mu_{NGC1559}= 31.388\pm0.051$~mag, and $\mu_{M101}= 29.070\pm0.043$~mag. These distances are in very good agreement with previous results in the literature based on Cepheids and TRGB \citep{riess2022, scolnic2023}. 

In the second step, these apparent zero-points of Mira PLRs in SNe host galaxies were compared with the standardized apparent SN Ia magnitudes, $m^0_B=12.141\pm0.086$ mag for SN 2005df in NGC 1559 \citep{scolnic2022} and   $m^0_B=9.808\pm0.116$ mag for SN 2011fe in M101 \citep{scolnic2023}. Similar to \citet{scolnic2023, huang2024}, the difference in the apparent magnitudes of the Miras and SN Ia in the target galaxies is defined as:
\begin{equation}
\Delta S = a_{targets} - m^0_B. \nonumber
\end{equation}

From the two $\Delta S$ measurements in the target galaxies ($12.957\pm0.093$~mag in NGC 1559 and $12.972\pm0.119$~mag in M101, we take a weighted average ($\overline{\Delta S}$) given by inverse-variance weighting. This average difference between the magnitude of a Mira with $\log P=2.3$ and the peak apparent magnitude of an SN Ia is connected to the fiducial peak luminosity of SN Ia by:
\begin{equation}
M^0_B = a_{0} - \overline{\Delta S}. 
\end{equation}

From a $\overline{\Delta S}=12.963\pm0.073$, we obtained $M^0_B =-19.253\pm0.079$ mag. The fiducial peak luminosity of SN Ia is in excellent agreement with the value independently calibrated using classical Cepheids in \citet[$-19.253$~mag,][]{riess2022} and differs by 0.02 mag from the measurement of \citet{huang2024} based on Mira calibrations.

\subsection{Hubble flow, {$H_0$}, and systematics}

\begin{deluxetable}{llccc}
\tablecaption{Error budget on $H_0$. \label{tbl:err_h0}}
\tablewidth{0pt}
\tablehead{
{Term} & {Description} & \multicolumn{3}{c}{$\sigma$ (mag)}\\
        &               &       Cluster     & LMC   & NGC 4258}
\startdata
$\sigma_{\mu_\textrm{anchor}}$   & Anchor distance   & 0.029 &   0.026   & 0.0324\\
$\sigma_\textrm{PL, anchor}$   & Mean of PLR in anchor   & 0.020 &   0.010   & 0.025\\
$\sigma_\textrm{[Fe/H]}$   & Anchor-host metallicity   & 0.036 &   0.045   & 0.018\\
 & Individual anchor total &  0.050 &  0.053 & 0.045 \\
 \vspace{-20pt}\\
    &       &   \multicolumn{2}{c}{$\underbrace{~~~~~~~~~~~~~~~~~~~~~~}$} & $\underbrace{~~~~~~~~~~~~}$\\
    &        &  \multicolumn{2}{c}{0.036} & 0.045 \\
$\sigma_\textrm{H,HST}$   & $H$ to F160W transformation   & \multicolumn{2}{c}{0.02} & 0.00\\ 
\hline
\multicolumn{2}{c}{All anchor subtotal} &  \multicolumn{3}{c}{0.030}\\
\hline
$\sigma_{\textrm{PL}/\sqrt{n}}$   &    Mean of PL in SNe hosts  &   \multicolumn{3}{c}{0.019}\\
$\sigma_{\textrm{SN}/\sqrt{n}}$   &    Mean of SNe calibrators ($\#$ SN)  &   \multicolumn{3}{c}{0.069 (2)}\\
$\sigma_{m-z}$    &   SN Ia $m-z$ relation    &   \multicolumn{3}{c}{0.0085}\\
$\sigma_\textrm{PL, slope}$    &   PLR slope, $\Delta \log P$, anchor-host    &   \multicolumn{3}{c}{0.01}\\
$\sigma_\textrm{extinction}$    &   Differential extinction, anchor-host    &   \multicolumn{3}{c}{0.02}\\
\hline
\multicolumn{2}{c}{Total uncertainty} &  \multicolumn{3}{c}{$0.081 \sim 3.7\%$ \textrm{on $H_0$}}
\enddata
\tablecomments{All terms represent close approximation of error values propagated in different rungs of the Mira distance scale. For the individual anchor total, 0.036 mag is the weighted variance of systematics for the Cluster and LMC anchors.}
\end{deluxetable}

The third rung of the distance ladder is independent of primary calibrators like Miras, and consists of SNe Ia in the Hubble flow, where the motion of galaxies is dominated by the expansion of the Universe. Using a set of SNe Ia that measure the expansion rate, \citet{riess2022} obtained the intercept $a_B$ of their distance (or magnitude) -- redshift relation. In the low-redshift limit ($z\approx0$), this is simply determined from $a_B = \log cz - 0.2 m^0_B$. \citet{riess2022} provided a baseline value of $a_B=0.7142\pm0.0017$\footnote{In Table 3 of \citet{huang2024}, the two uncertainties on the intercept of SN Ia Hubble diagram correspond to $a_B$ ($\sigma=0.000176$) and $5a_B$ ($\sigma=0.0085$), respectively.} using 277 SNe Ia in the Hubble flow with $0.0233 < z < 0.15$. This $a_B$ value was determined including the two SNe Ia hosts with known Miras in our sample, and thus is suitable also for our late-type galaxies containing both Cepheids and Mira variables.

Finally, the $H_0$ solution is given by:
\begin{equation}
\log H_0 = (M^0_B +5a_B +25)/5, 
\end{equation}

\noindent where $M^0_B$ is the calibrated fiducial peak luminosity of SNe Ia determined in Section~\ref{sec:sne_dis}.  We obtained an  $H_0 = 73.06\pm2.67$ km~s$^{-1}$~Mpc$^{-1}$, a $3.7\%$ determination of the Hubble constant including both statistical and systematic uncertainties.

Table~\ref{tbl:err_h0} provides a detailed breakdown of the error budget on the $H_0$. The total uncertainty is dominated by the $\sim2.3\%$ statistical uncertainty on the fiducial peak luminosity of SNe Ia in two host galaxies. With a global slope that is determined using Miras in three-anchor galaxies, we obtained an improved constraint on the anchor zero-point luminosity of a Mira with $\sim200$ d period. The absolute value of the anchor zero-point is similar to \citet{huang2024} value, but the uncertainty is reduced significantly considering that the metallicity systematics are also included in the anchor zero-point. The metallicity systematic is propagated through the coefficient of $0.09\pm0.04$ with a plausible difference in the mean metallicity of $\pm0.2/0.4/0.5$ dex between NGC 4258/Cluster/LMC and SNe Ia host galaxies. We also include a systematic error of anchor distances and also add 0.02 mag for the $H$ to F160W transformations for the ground-based LMC and Cluster calibrations. The statistical uncertainties due to SNe Ia host galaxies and the slope of the PLR are similar to those from \citet{huang2024}. The uncertainty due to interstellar extinction in SNe Ia host galaxies is propagated assuming that the reddening is similar to the LMC, where a difference of 0.02 mag is obtained with and without interstellar reddening correction. The global slope together with improved anchor zero-points provides better constraints on the ${\Delta S}$ for individual SNe Ia host galaxies. The difference
in ${\Delta S}$ of $\sim0.015$ mag between two SNe host galaxies suggests excellent consistency in the fiducial peak luminosity of their SNe Ia. Finally, the inverse-weighted average value of fiducial peak luminosity of SNe Ia is better constrained with an error $14\%$ smaller than the estimated uncertainty in \citet{huang2024}. These small improvements resulting from an additional anchor in the $H_0$ solution provide a significant reduction in the error budget to $3.7\%$ in the Mira distance ladder.

\section{Discussion}
\label{sec:discuss}

Homogeneous photometry of long-period variables at infrared wavelengths is important to calibrate their luminosity scales at high precision for complementing or testing Cepheid based distances to SNe Ia host galaxies. We utilized time-series observations of 41 O-rich Miras in globular clusters to calibrate their absolute luminosity scales in near-infrared wavelengths. The time-series photometric data together with accurate distances to globular clusters from \citet{baumgardt2021} result in PLRs that are in good agreement with O-rich Miras in the LMC. The scatter in these relations is significantly small and is comparable to Cepheids in the LMC. The purely geometric calibration based on mean-parallaxes of the globular cluster from \citet{vasiliev2021} further supports the accuracy of these PLR calibrations. However, 14 Miras in our sample are beyond 10 kpc distance and therefore, large parallax uncertainties lead to a larger scatter in these relations as compared to those derived using distances from  \citet{baumgardt2021}. The improvements in the parallaxes of globular clusters with the next Gaia data release may provide a comparable accuracy and improved precision for the Mira based calibrations. Nevertheless, these absolute luminosity calibrations of cluster Mira variables provide a crucial anchor to complement LMC and NGC 4258. Similar to classical Cepheids, these three anchors form a baseline solution for the first step of the distance ladder used in the local Hubble constant determinations. 

The O-rich Mira candidates in our sample were selected as the most likely members of their host clusters, which cover a wide range of metallicity $-1.7<\textrm{[Fe/H]}<-0.1$. This offers a unique opportunity to empirically quantify possible dependence of Mira PLRs on the iron abundance. As an additional parameter in the PLRs, we found no statistically significant contribution of metallicity on Mira luminosities in globular clusters. The metallicity coefficients were found to be of the order of $\lesssim -0.002\pm0.002$ in all three $JHK_s$ bands, empirically supporting previous results of null or small metallicity dependence \citep{whitelock2008, bhardwaj2019, goldman2019}. \citet{rejkuba2004} also found that the $K_s$ band PL relation for Miras in NGC 5128, which has a photometric metallicity of $-0.25$~dex, is similar to that of LMC suggesting no significant effect of metallicity on distance determination. If this result holds, it will be an important advantage for Mira distance scale as classical Cepheids exhibit a clear dependence on metallicity \citep{breuval2022, bhardwaj2023a, trentin2024}. However, the majority of clusters with Miras have a small metallicity range from $-0.3$ to $-0.7$~dex and therefore it may not be possible to quantify this dependence accurately based on our small sample of variables. 

Another approach to quantifying metallicity effects involves comparing the calibrated zero-points of the PLRs in anchor galaxies with different metallicities. Using cluster, LMC, and NGC 4258 Miras, we find a small slope ($0.09\pm.04$ mag/dex) of the variation in the zero-points as a function of metallicity. This determination of metallicity effects is also very sensitive to the uncertainties on distances to these systems, as well as the adopted mean metallicities for their Mira stars. For the distant systems observed with {\it HST}, we assume the mean metallicity based on H$\texttt{II}$ regions that is used in the Cepheid distance ladder \citep{riess2022a}. This approximation may not be appropriate for intermediate-age Mira populations, and therefore, we do not include the metallicity term in the distance scale analysis. However, these mean metallicity values can be used to estimate the relative difference in abundances between anchor and host galaxies, assuming that it is similar for young and intermediate-age populations. This allowed us to provide a reasonable estimate of systematic uncertainty due to metallicity in the $H_0$ determinations. If we propagate a metallicity term of $0.09\pm.04$ mag/dex in the Mira distance ladder adopting the metallicity of SNe Ia host galaxies from \citet{riess2022}, we find a small difference of $<0.2$ units in $H_0$. Homogeneous spectroscopic metallicity determinations of intermediate-age stars or AGB stars in these galaxies will be useful to precisely quantify the metallicity effects based on the zero-points of anchor PLRs. 

One of the limitations for the use of Miras for the distance scale is the poor understanding of circumstellar extinction in their extended atmospheres. However, infrared observations help mitigate these uncertainties due to smaller extinction effects at longer wavelengths. While short-period O-rich Miras are less sensitive to these effects, including extinction corrections to derive their PLRs is not straightforward for the variables in distant galaxies. For the cluster Miras, we explored different variants by changing the adopted reddening law or the assumption of total-to-selective absorption ratios. We found that the coefficients of PLRs remain unchanged within $1-2\sigma$ of their quoted uncertainties. For the LMC, we found a variation of up to 0.02 mag in the PLR zero-points with or without the correction for interstellar extinction, which is added as uncertainty in the $H_0$ determination. For the distant galaxies observed with {\it HST}, only MW foreground extinction corrections are applied neglecting interstellar reddening. These assumptions together with metallicity effects on PLRs can lead to a possible systematic uncertainty of 0.045 mag in the error budget for $H_0$ \citep{huang2024}. 

Finally, the Mira based $H_0 = 73.06\pm2.67$ km~s$^{-1}$~Mpc$^{-1}$ derived in this work provides an independent comparison to Cepheid and TRGB based determinations. As previously mentioned, the uncertainty on the $H_0$ value primarily comes from the peak fiducial luminosity of SNe Ia, and additional SNe Ia hosts with Mira variables are required to minimize these statistical uncertainties. Nevertheless, our $3.7\%$ accurate determination of $H_0$ is in excellent agreement with the three-anchor baseline solution from Cepheid-based local measurements. The $H_0$ value based on intermediate-age Mira stars is also in agreement with Population II TRGB-based measurements from \citet{freedman2024}. Higher-precision Mira-based $H_0$ values may provide insight into possible systematics due to stellar standard candles of different age and metallicity populations. Currently, our $H_0$ estimates are in excellent agreement with Cepheid-based values within $0.1\sigma$, and further strengthen the tension with early universe measurements of the expansion rate of the Universe.

\begin{acknowledgments}

A.B. thanks the funding from the Anusandhan National Research Foundation (ANRF) under the Prime Minister Early Career Research Grant scheme (ANRF/ECRG/2024/000675/PMS). This research was supported by the International Space Science Institute (ISSI) in Bern/Beijing through ISSI/ISSI-BJ International Team project ID $\#$24-603 – ``EXPANDING Universe'' (EXploiting Precision AstroNomical Distance INdicators in the Gaia Universe). C.D.H. is supported by the National Science Foundation Astronomy \& Astrophysics Postdoctoral Fellowship under Grant No. 2401770.
This research was supported by the Munich Institute for Astro-, Particle and BioPhysics (MIAPbP) which is funded by the Deutsche Forschungsgemeinschaft (DFG, German Research Foundation) under Germany´s Excellence Strategy – EXC-2094 – 390783311.
\end{acknowledgments}

\begin{contribution}

AB developed this research project and was responsible for the data analysis and writing the manuscript. NM provided datasets of Mira variables in globular clusters. CDH provided Mira data in NGC 4258 and NGC 1559 that have been recalibrated and zero-point corrected to a consistent magnitude system. AGR and MR contributed to the formal analysis and validation, and all authors edited the final version of the manuscript. 

\end{contribution}

\facilities{SAAO (SIRIUS/IRSF)}

\software{\texttt{The IDL Astronomy User's Library} \citep{landsman1993}, \texttt{Astropy} \citep{astropy2013, astropy2018, astropy2022}}

\appendix

\section{Light curves of Long-Period Variables in globular clusters}
\label{sec:supp_fig}

\begin{figure*}[ht!]
\plotone{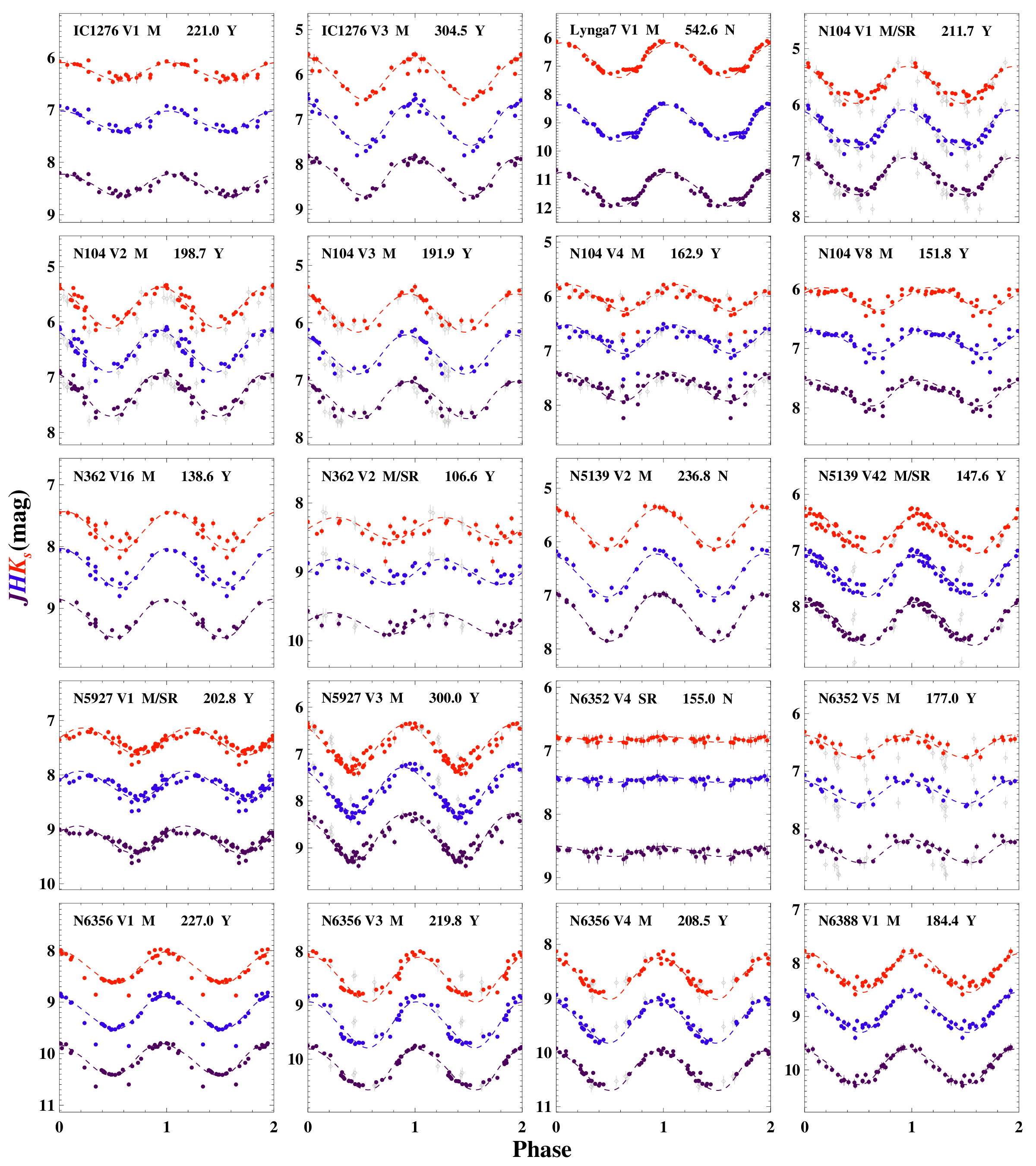}
\caption{Template-fitted light curves of 55 long-period variables analyzed in this work in $J$ (violet), $H$ (blue), and $K_s$ (red). The light curves in $H/K_s$ bands in blue/red are offset by $-0.1/-0.5$~mag for visualization purposes. The grey data points represent literature photometry used to extend the time baseline. Cluster variable ID, period, variable type, and the PLR flag (Y/N) are shown on top of each panel.}
\label{fig:lpv_lc}
\end{figure*}

\begin{figure*}[ht!]
\plotone{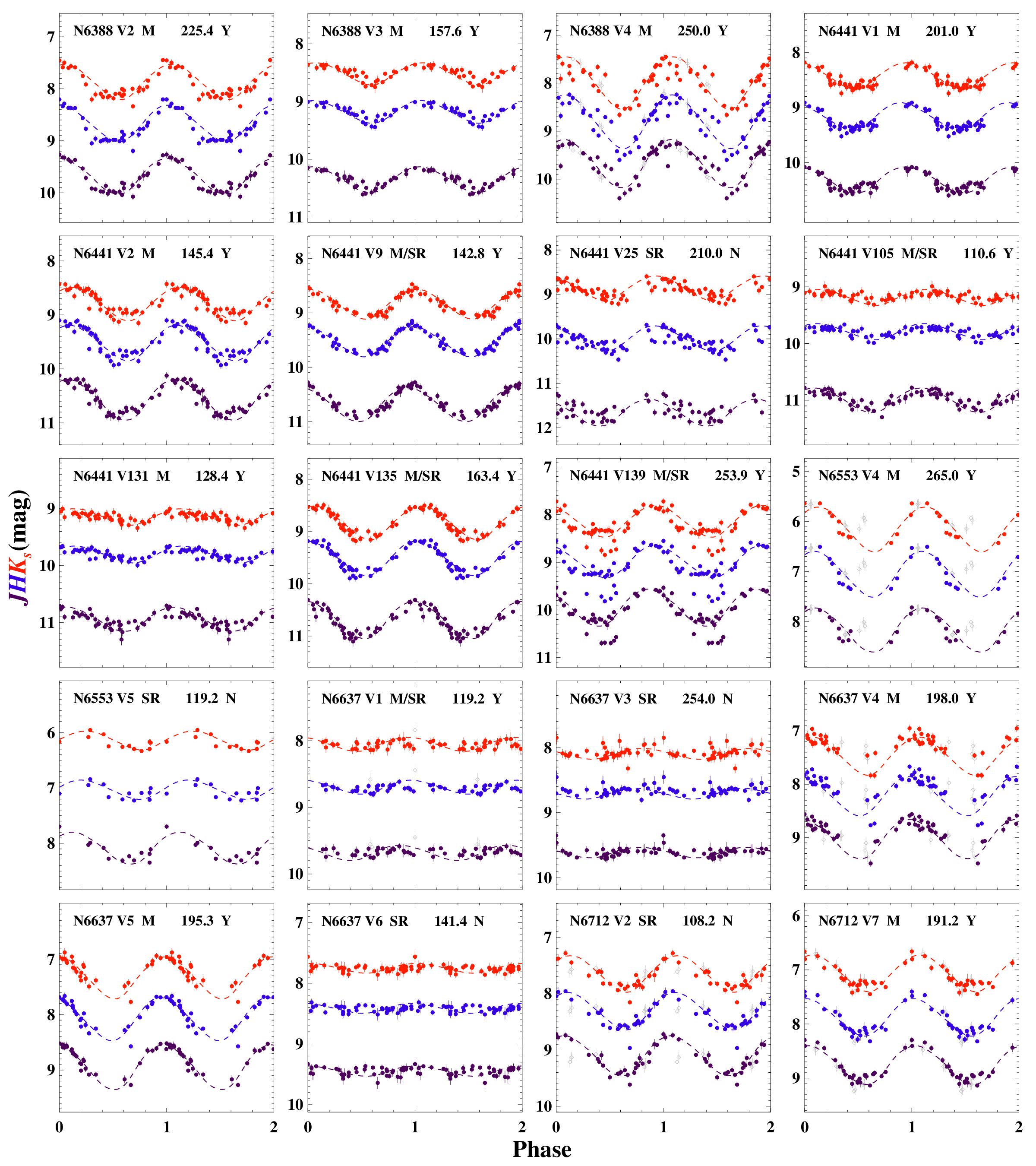}
\caption{Fig.~\ref{fig:lpv_lc} continued.}.
\label{fig:lpv_lc1}
\end{figure*}

\begin{figure*}[ht!]
\plotone{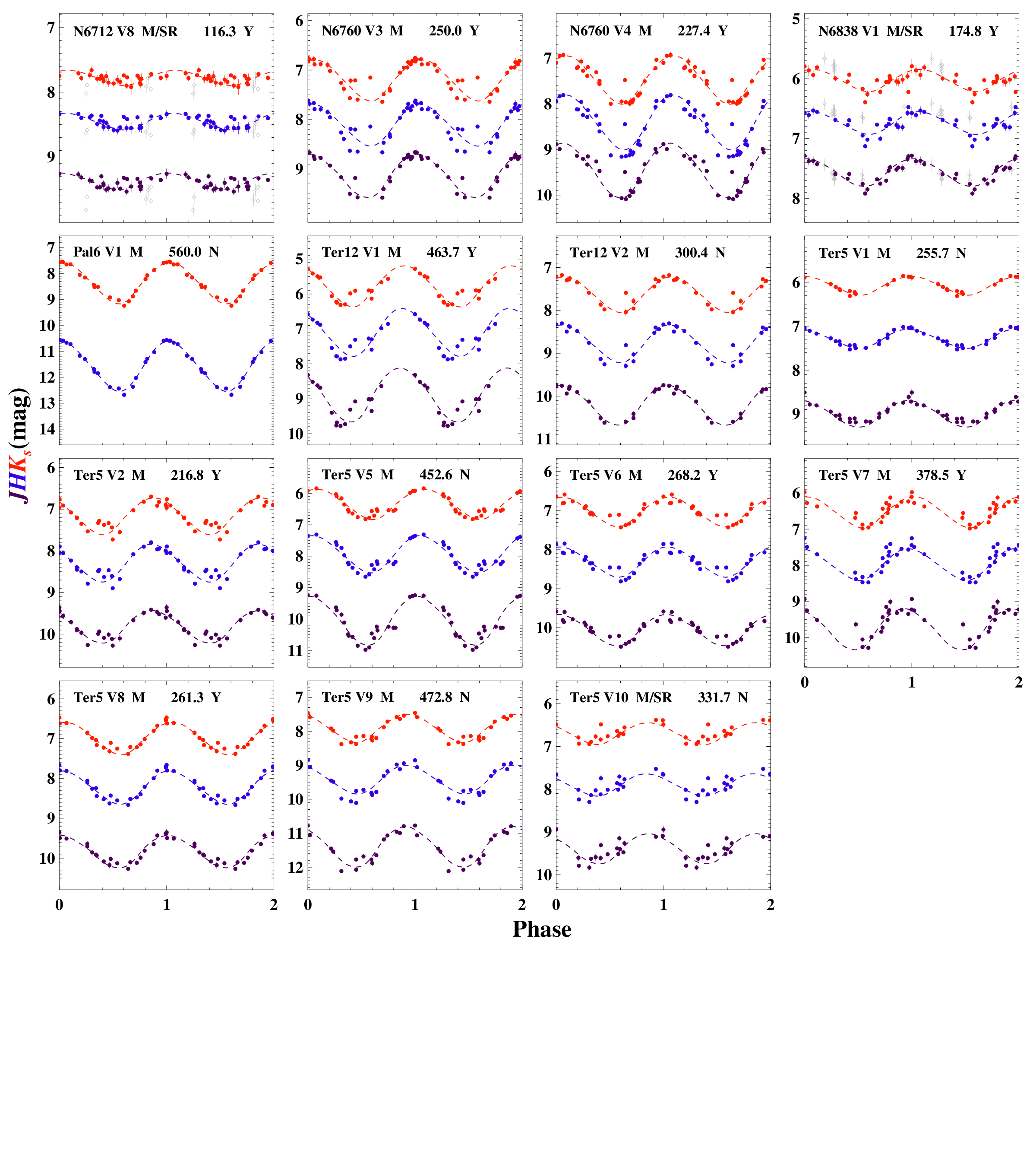}
\vspace{-90pt}
\caption{Fig.~\ref{fig:lpv_lc} continued.}.
\label{fig:lpv_lc2}
\end{figure*}

Figs~\ref{fig:lpv_lc}-\ref{fig:lpv_lc2} display template-fitted IRSF light curves of all 55 LPVs used in this work. 

\section{Photometric data for the field Miras in the Milky Way}
\label{sec:supp_tbl}

Table~\ref{tbl:lpv_fielddata} provides pulsation properties and NIR photometry of 40 field Miras in the Milky Way taken from \citet{whitelock2000} and \citet{whitelock2008}. 

\begin{deluxetable}{cccccccccccccc}[!h]
\tablecaption{Pulsation properties of 40 field Miras in the Milky Way. \label{tbl:lpv_fielddata}}
\tabletypesize{\footnotesize}
\tablehead{
{Star ID} & {Gaia DR3 ID} & {$P$ (days)} & {Class} & \multicolumn{3}{c}{magnitudes} & \multicolumn{3}{c}{$\sigma_\textrm{mag}$} & \multicolumn{3}{c}{amplitudes} & $E(B-V)$ \\
      &      &       &       &   $J$ &   $H$ &   $K_s$ & $J$ &   $H$ &   $K_s$ & $J$ &   $H$ &   $K_s$ &
}
\startdata
       BQ Pav& 6423407784861791872&    109.5&      M&     7.65&     6.81&     6.52&     0.03&     0.03&     0.03&     0.68&     0.58&     0.56&     0.05\\
       RR Sgr& 6750288640427640192&    336.4&      M&     1.91&     1.03&     0.61&     0.03&     0.04&     0.04&     0.81&     0.79&     0.77&     0.13\\
       RT Aqr& 6628502987323579904&    246.5&      M&     3.47&     2.51&     2.17&     0.02&     0.02&     0.03&     0.25&     0.20&     0.20&     0.03\\
       RT Cen& 6115711861009879296&    268.0&      M&     4.01&     3.06&     2.73&     0.02&     0.02&     0.03&     0.35&     0.34&     0.33&     0.07\\
       RU Hya& 6269317277540296064&    330.5&      M&     3.03&     2.19&     1.74&     0.03&     0.03&     0.03&     0.54&     0.54&     0.53&     0.06\\
       RU Lib& 6264954243602863360&    323.8&      M&     3.56&     2.65&     2.27&     0.04&     0.03&     0.03&     0.88&     0.72&     0.70&     0.12\\
       RU Oct& 4616474464582858368&    387.6&      M&     4.12&     3.07&     2.67&     0.02&     0.02&     0.03&     0.24&     0.27&     0.26&     0.09\\
       RV Pup& 5557401915377587328&    188.6&      M&     4.70&     3.79&     3.50&     0.03&     0.03&     0.04&     0.68&     0.74&     0.72&     0.10\\
       RV Sgr& 4044906332916568192&    315.1&      M&     2.97&     2.01&     1.63&     0.03&     0.03&     0.03&     0.48&     0.47&     0.45&     0.21\\
       RX Cen& 6115672347311117184&    335.2&      M&     4.32&     3.40&     2.90&     0.04&     0.03&     0.03&     0.94&     0.71&     0.69&     0.08\\
       RX Tau& 3292447407136388480&    324.3&      M&     2.74&     1.72&     1.20&     0.03&     0.03&     0.03&     0.77&     0.55&     0.54&     0.19\\
       RZ Sco& 6236076189188130432&    156.6&      M&     5.31&     4.48&     4.11&     0.03&     0.03&     0.03&     0.63&     0.59&     0.57&     0.11\\
        R Crv& 3518127631936756736&    315.6&      M&     3.13&     2.27&     1.87&     0.03&     0.03&     0.03&     0.71&     0.65&     0.64&     0.06\\
        R Peg& 2715274995932228352&    383.3&      M&     1.86&     0.95&     0.42&     0.04&     0.03&     0.03&     0.89&     0.57&     0.56&     0.05\\
        R Ret& 4675572149422764160&    278.8&      M&     3.06&     2.20&     1.77&     0.03&     0.03&     0.03&     0.77&     0.69&     0.68&     0.03\\
        R Ser& 1192855977386610688&    359.1&      M&     1.98&     1.07&     0.59&     0.04&     0.04&     0.04&     0.91&     0.90&     0.88&     0.04\\
        R Sgr& 4083847117722733440&    268.6&      M&     3.29&     2.39&     2.02&     0.03&     0.03&     0.04&     0.79&     0.77&     0.75&     0.12\\
        R Vir& 3709971554622524800&    144.6&      M&     3.21&     2.39&     2.03&     0.03&     0.03&     0.03&     0.60&     0.60&     0.59&     0.02\\
        S CMi& 3143124657116728448&    339.0&      M&     1.76&     0.85&     0.42&     0.03&     0.03&     0.03&     0.58&     0.50&     0.49&     0.06\\
        S Hya& 578095970907854976&     258.2&      M&     4.10&     3.22&     2.86&     0.03&     0.03&     0.03&     0.51&     0.61&     0.60&     0.04\\
        S Peg& 2761866801159498368&    316.0&      M&     2.79&     1.85&     1.44&     0.03&     0.03&     0.03&     0.65&     0.53&     0.51&     0.05\\
        S Scl& 2316495425756704384&    378.9&      M&     1.72&     0.83&     0.41&     0.03&     0.03&     0.03&     0.71&     0.62&     0.61&     0.01\\
        S Sex& 3806792486979461760&    266.7&      M&     4.73&     3.79&     3.44&     0.03&     0.03&     0.03&     0.59&     0.55&     0.53&     0.07\\
        T Aqr& 6913517223245165696&    200.3&      M&     4.54&     3.65&     3.28&     0.03&     0.03&     0.03&     0.58&     0.45&     0.43&     0.05\\
        T Cap& 6836403658129756416&    276.5&      M&     4.57&     3.68&     3.25&     0.04&     0.03&     0.03&     0.99&     0.64&     0.63&     0.07\\
        T Col& 4826242419666284800&    225.1&      M&     3.15&     2.28&     1.92&     0.03&     0.03&     0.03&     0.78&     0.68&     0.67&     0.02\\
        T Eri& 5084084630886477952&    250.9&      M&     3.72&     2.82&     2.44&     0.03&     0.03&     0.03&     0.64&     0.61&     0.60&     0.04\\
        T Hor& 4748535741042603520&    220.3&      M&     4.53&     3.64&     3.30&     0.03&     0.03&     0.03&     0.53&     0.45&     0.44&     0.02\\
        T Hya& 5749870429386271488&    287.2&      M&     3.60&     2.70&     2.35&     0.02&     0.03&     0.03&     0.42&     0.43&     0.42&     0.04\\
        U Cet& 5170512944979310208&    243.0&      M&     3.88&     3.03&     2.66&     0.03&     0.04&     0.04&     0.78&     0.78&     0.76&     0.02\\
        W Cen& 5341387531702429824&    206.5&      M&     3.23&     2.32&     1.92&     0.03&     0.03&     0.03&     0.67&     0.59&     0.57&     0.86\\
        W Psc& 307883012627596288&     186.9&      M&     7.16&     6.26&     5.90&     0.03&     0.03&     0.03&     0.66&     0.64&     0.62&     0.06\\
        W Pup& 5535613889889944960&    119.8&      M&     4.66&     3.81&     3.47&     0.03&     0.03&     0.03&     0.60&     0.60&     0.58&     0.33\\
        W Vel& 5355341021422598272&    389.9&      M&     1.88&     0.92&     0.40&     0.05&     0.04&     0.04&     1.19&     0.95&     0.92&     1.02\\
        X Aqr& 6820838735304183424&    312.6&      M&     4.27&     3.43&     3.02&     0.04&     0.04&     0.04&     0.92&     0.84&     0.82&     0.03\\
        X Cen& 5380488432883509888&    341.7&      M&     2.36&     1.52&     1.01&     0.04&     0.03&     0.03&     0.82&     0.63&     0.61&     0.13\\
        X Cet& 3262686651029849216&    173.6&      M&     5.42&     4.48&     4.19&     0.03&     0.03&     0.03&     0.46&     0.48&     0.47&     0.06\\
        X Mon& 3050679330064623744&    151.0&      M&     3.97&     3.14&     2.85&     0.03&     0.03&     0.03&     0.56&     0.47&     0.46&     1.17\\
        Y Eri& 4718325284680111488&    312.7&      M&     3.11&     2.12&     1.73&     0.03&     0.03&     0.03&     0.45&     0.36&     0.35&     0.03\\
        Z Sco& 6243325922174300288&    360.1&      M&     2.80&     1.80&     1.48&     0.02&     0.03&     0.03&     0.34&     0.34&     0.33&     0.19\\
\enddata
\tablecomments{NIR mean-magnitudes and amplitudes are obtained using templates as discussed in Section~\ref{sec:subdata1}. The Gaia DR3 Source ID are provided in the second column. $E(B-V)$ values are taken from \citet{schlegel1998} reddening maps.}
\end{deluxetable}

\bibliography{mybib_final}{}
\bibliographystyle{aasjournalv7}



\end{document}